\documentclass[reprint,aps,prd,superscriptaddress,nofootinbib,floats,showpacs]{revtex4-2}

\usepackage{hyperref}
\usepackage{graphicx}
\usepackage{subfigure}
\usepackage{mathtools}
\usepackage{amsmath,amssymb,amsfonts,amsthm,latexsym,stmaryrd,physics,mathrsfs}
\usepackage{color}
\usepackage{xcolor}
\usepackage{bm}
\usepackage{geometry}
\usepackage{float}
\usepackage{nicefrac}
\usepackage{microtype}
\usepackage{verbatim}
\usepackage[T1]{fontenc}
\usepackage[utf8]{inputenc}
\usepackage{dsfont}
\usepackage{tikz}

\def\beq{\begin{equation}}
	\def\eeq{\end{equation}}
\newcommand{\bea}{\begin{eqnarray}}
	\newcommand{\eea}{\end{eqnarray}}
\def\bi{\begin{itemize}}
	\def\ei{\end{itemize}}
\def\bfig{\begin{figure}}
	\def\efig{\end{figure}}
\def\be{\begin{eqnarray}}
	\def\ee{\end{eqnarray}}
	
\newtheorem{theorem}{Theorem}[section]

\newtheorem{corollary}{Corollary}[theorem]
\newtheorem{lemma}[theorem]{Lemma}

\begin{document}
	
\title{Effective Dynamics of Loop Quantum Kaluza-Klein Cosmology}
	
\author{Shengzhi Li}
\affiliation{School of Physics and Astronomy, Key Laboratory of Multiscale Spin Physics (Ministry of Education), Beijing Normal University, Beijing 100875, China}

\author{Yongge Ma}
\email{Contact author: mayg@bnu.edu.cn}
\affiliation{School of Physics and Astronomy, Key Laboratory of Multiscale Spin Physics (Ministry of Education), Beijing Normal University, Beijing 100875, China}
	
\author{Faqiang Yuan}
\affiliation{School of Physics and Astronomy, Key Laboratory of Multiscale Spin Physics (Ministry of Education), Beijing Normal University, Beijing 100875, China}
	
\author{Xiangdong Zhang}
\email{Contact author: scxdzhang@scut.edu.cn}
\affiliation{School of Physics and Optoelectronics, South China University of Technology, Guangzhou 510641, China}

\begin{abstract}
The five-dimensional loop quantum Kaluza-Klein cosmology is constructed based on the symmetric reduction of the connection formulation of the full theory. Through semiclassical analysis, the effective scalar constraint for the cosmological model coupled with a dust field is derived, incorporating the quantum fluctuations of geometry as a subleading order correction. It demonstrates that the quantum model has the correct classical limit. The explicit solutions to the equations of motion show that the big bang and past big rip singularities in the classical model are avoided by a quantum bounce and a quantum collapse respectively in the effective model. In a particular scenario, the dynamical compactification of the extra dimension is realized, while the observable four-dimensional universe transitions through three distinct epochs: (i) a super-inflationary phase generating 55 e-folds, (ii) a decelerated expansion era, and (iii) a late-time accelerated expansion phase driven by quantum fluctuations. These results suggest that both cosmic inflation and dark energy may originate from the interplay between the compact extra dimension and quantum geometric effects.
\end{abstract}

\maketitle 

\noindent\textit{Introduction.} While classical cosmology based on general relativity has achieved remarkable success in describing cosmic evolution, it confronts fundamental limitations in addressing pivotal challenges including the big bang singularity problem, the inflationary dilemma, and the origin of dark energy  \cite{Penrose:1964wq,Hawking:1973uf,Guth:1980zm,Linde:1981mu,Albrecht:1982wi,SupernovaSearchTeam:1998fmf,SupernovaCosmologyProject:1998vns}. It is naturally expected to address these issues by certain quantum-gravitational extension of the classical framework. However, a natural mechanism generates inflation together with late-time cosmic acceleration is still missing. From the theoretical viewpoint, the inconsistency between classical general relativity and quantum mechanics indicates the necessity of a quantum theory of gravity.  On one hand, among various approaches to quantum gravity, loop quantum gravity (LQG) is notable for its advantage of background independence \cite{A. Ashtekar and Y. Ma,Ashtekar:2004eh,Rovelli:2004tv,Thiemann:2007pyv,Han:2005km}. On the other hand, the idea of unifying gravity with other fundamental interactions of matter fields by introducing extra dimensions of spacetime is proposed in the well-known Kaluza-Klein (KK) theory \cite{Kaluza:1921tu,Klein:1926tv,bailin1987kaluza,Appelguist}. Thus, it is desirable to extend LQG  to higher dimensional gravity \cite{Bodendorfer:2011nv,Bodendorfer:2011nx,Long:2019nkf,Long:2020wuj,Long:2022thb}, which might lead to a quantum geometrical framework to unify gravity and matter.

As the cosmological models of LQG, loop quantum cosmology (LQC) provides a compelling resolution to the big bang singularity through a quantum bounce mechanism \cite{Ashtekar:2003hd,Ashtekar:2011ni,Ashtekar:2006rx,Ashtekar:2006wn,Ding:2008tq,Yang:2009fp,Zhang:2021zfp}. It is also argued in some models of LQC that the quantum geometry corrections of LQC may even persist at macroscopic scales \cite{Ding:2008tq,Yang:2009fp,Zhang:2021zfp}. While a consistent resolution of all the above issues of classical cosmology by four-dimensional (4D) LQC has not been achieved, in this letter we are going to demonstrate this possibility by combining the ideas of LQG and KK theory within a novel ‌5D KK LQC framework, based on the symmetric reduction of the connection formulation of the full theory for the first time‌. One of our models achieves two transformative results. First, the dynamical compactification of the fifth dimension naturally generates the expansion of the observed 4D spacetime \cite{Mohammedi:2002tv,Qiang:2004gg}. Second, the post-bounce evolution in the visible dimensions spontaneously initiates a super-inflationary epoch \cite{Ashtekar:2011ni}, bypassing the fine-tuning issue of the conventional mechanisms of inflation, followed by a late time deceleration to acceleration transition driven by quantum fluctuations. This dual mechanism bridges early- and late-universe dynamics within a unified quantum-geometric paradigm.

\vspace{\baselineskip}
\noindent\textit{Classical setting.} Let us consider the homogeneous 5D KK cosmology with spatial manifold $\Sigma^{4}$ of topology $\mathbb{R}^3\!\times\!\mathbb{S}^1$ and the corresponding spatial isometry group $\mathrm{E}(3)\!\times\!\mathrm{U}(1)$. Denote $(x^{i},y)$ with $i=1,2,3$ as the natural coordinates of $\mathbb{R}^3\!\times\!\mathbb{S}^1$. As usual, we fix a fiducial cell $\Sigma^3_{(0)}\!\equiv\!(0,V_0^{1/3})^3\subset\!\mathbb{R}^3$ and restrict the phase space variables to $\Sigma^3_{(0)}\!\times\!\mathbb{S}^1$ to avoid integral divergences. In the geometric sector of 5D LQG \cite{Bodendorfer:2011nv}, the canonical variables consist of the $\mathrm{Spin}(5)$ connections $A_{aIJ}$ and their conjugate momenta $\pi^{bKL}$ defined on $\Sigma^{4}$. Based on the underlying symmetries, they take the form (see Appendix \ref{appendixA} for details):
\begin{eqnarray}
A_{aIJ}\frac{L^{IJ}}{2}dx^a&=&\frac{A_{1}}{V_{0}^{\frac{1}{3}}}L_{i5}dx^i+A_{y}L_{45}dy,\\
\pi^{aIJ}\frac{L_{IJ}}{2}\partial_a&=&\frac{\pi^{1}}{V_{0}^{\frac{2}{3}}}L^{i5}\partial_i+\frac{\pi^{y}}{V_{0}}L^{45}\partial_y,
\end{eqnarray}
where $L_{IJ}$ are the generators of the $\mathrm{Spin}(5)$ group. Thus, $(A_1,\pi^{1})$ and $(A_y,\pi^{y})$ form the phase space for the geometric sector of KK cosmology. They satisfy the reduced non-trivial Poisson brackets: $\{A_1,\pi^{1}\}=\beta\kappa/3$, $\{A_y,\pi^{y}\}=\beta\kappa$, where $\kappa\equiv8G^{(5)}c^{-3}$ with $G^{(5)}$ and $c$ being the 5D gravitational constant and the speed of light respectively, and $\beta$ is the 5D Immirzi parameter \cite{Bodendorfer:2011nv}. 

It is convenient to introduce the following new conjugate variables via the canonical transformation:
\begin{align}
	\bar{A}_1&\coloneq \bar{\mu}_{1}A_1,          &\bar{\pi}^{1}&\coloneq\frac{\pi^1}{\bar{\mu}_{1}}, \label{pi1}\\
	\bar{A}_y&\coloneq A_y - A_1\frac{\pi^{1}}{\pi^{y}},  &\pi^y&\coloneq\pi^{y},\label{piy}
\end{align}
where $\bar{\mu}_{1}\equiv\Delta^{1/2}|\pi^{y}|^{-1/3}$ with  $\Delta^{3/2}\equiv2\pi\beta\kappa\hbar$ being the minimal 3D area in 5D LQG theory \cite{Bodendorfer:2011nx}. The momentum variables are related to the spacetime metric as $ds^2=-N^2dt^2+a^2\delta_{ij}dx^{i}dx^{j}+b^2dy^2$, where $a^2=(\pi^y/V_0)^{2/3}$, $b^2=\Delta(\bar{\pi}^1/\pi^y)^2$, and $N$ is the lapse function. Additionally, the dynamics of the system are encoded in the scalar constraint, with its geometric part given by:
\begin{equation}
C_{gr}[N]=\frac{-3N\Delta^{-\frac{1}{2}}}{2\beta^{2}\kappa|\bar{\pi}^{1}|}[2(\bar{A}_1\bar{\pi}^{1})^{2}+\bar{A}_1\bar{A}_{y}\bar{\pi}^{1}\pi^{y}].
\end{equation}

Since the present universe is dominated by the matter, except for dark energy, we also introdue the phase space of the matter sector using the dust variables $(T,P_{T})$ with the Poisson bracket $\{T,P_{T}\}=1$, where $T$ represents the comoving time of the dust field, and $P_{T}$ is the dust mass within the fiducial cell. Thus, the total scalar constraint turns out to be $C[N]=C_{gr}[N]+NcP_{T}$, which determines a universe with both the big bang and the past big rip singularities \cite{SFXY,OG}.

\vspace{\baselineskip}
\noindent\textit{Quantum dynamics and its semiclassical analysis.} In the KK LQC, a polymer-like quantization of the connection variables is implemented by mimicking the quantization of the holonomy-flux algebra proposed in the full theory (see e.g. Refs.\cite{Ashtekar:2004eh,Rovelli:2004tv,Thiemann:2007pyv,Han:2005km,Bodendorfer:2011nx}). It turns out that the common eigenstates $\{|\lambda,\xi\rangle\mid\lambda,\xi\!\in\!\mathbb{R}\}$ of the self-adjoint operators $\hat{\bar{\pi}}^{1}$ and $\hat{\pi}^{y}$, corresponding to the momentum variables in Eqs.(\ref{pi1}) and (\ref{piy}), form an orthonormal basis in the kinematic Hilbert space $\mathcal{H}_{gr}$ of the geometric sector \cite{Zhang:2015bxa}, such that $\langle\lambda',\xi'|\lambda,\xi\rangle=\delta_{\lambda',\lambda}\delta_{\xi',\xi}$, where the Kronecker-$\delta$ symbol is used. The actions of the basic operators on $|\lambda,\xi\rangle$ read: 
\begin{align}
\widehat{e^{i\frac{\bar{A}_{1}}{2}}}|\lambda,\xi\rangle\!&=\!|\lambda\!+\!1,\xi\rangle,
&\hat{\bar{\pi}}^{1}|\lambda,\xi\rangle\!&=\!\frac{\hbar\beta\kappa\lambda}{6}|\lambda,\xi\rangle,\\
\widehat{e^{i\frac{\bar{A}_{y}}{2}}}|\lambda,\xi\rangle\!&=\!|\lambda,\xi\!+\!1\rangle,
&\hat{\pi}^{y}|\lambda,\xi\rangle\!&=\!\frac{\hbar\beta\kappa\xi}{2}|\lambda,\xi\rangle.
\end{align}
In analogy with Refs.\cite{Ashtekar:2003hd,Ashtekar:2006rx,Ashtekar:2006wn}, by regularizing the curvature using holonomies around square loops, the scalar constraint $C_{gr}[N]$ can be promoted to the operator 
\begin{align}
\hat{C}_{gr}[N]\!=\!\frac{N\hbar\beta^{-1}}{16\Delta^{\frac{1}{2}}}(\sum_{r=0,\pm4}\hat{u}_{r}\!+\!\sum_{k,l=\pm2}\hat{u}_{k,l}),
\end{align}
such that
\begin{align}
	\hat{u}_{0}|\lambda,\xi\rangle\!&\coloneq\!-4\lambda^{2}\eta^{2}(\lambda)|\lambda,\xi\rangle,\nonumber\\
	\hat{u}_{\pm4}|\lambda,\xi\rangle\!&\coloneq\!2\eta(\lambda\!\pm\!4)\lambda^{2}\eta(\lambda)|\lambda\!\pm\!4,\xi\rangle,\nonumber\\
	\hat{u}_{\pm2,2}|\lambda,\xi\rangle\!&\coloneq\!\pm\frac{3}{\mu_{y}}\eta(\lambda\!\pm\!2)\lambda\xi\eta(\lambda)|\lambda\!\pm\!2,\xi\!+\!2\mu_{y}\rangle,\nonumber\\
	\hat{u}_{\pm2,-2}|\lambda,\xi\rangle\!&\coloneq\!\mp\frac{3}{\mu_{y}}\eta(\lambda\!\pm\!2)\lambda\xi\eta(\lambda)|\lambda\!\pm\!2,\xi\!-\!2\mu_{y}\rangle,\nonumber
\end{align}
where, $\eta(\lambda)\!\equiv\!|\lambda|^{-1/2}$ if $\lambda\!\ne\!0$, and otherwise $\eta(\lambda)\!=\!0$. The regularization parameter $\mu_{y}$ in the above equations will be simply chosen to be $2\pi$, whose choices would not affect the physics results. Moreover, since $C_{gr}[N]$ is a real function, we use the natural self-adjoint extension $\hat{C}^{sym}_{gr}[N]\!\coloneq\!(\hat{C}_{gr}[N]\!+\!\hat{C}^{\dagger}_{gr}[N])/2$ as its quantum counterpart.

To ensure that $\hat{C}^{sym}_{gr}[N]$ is a viable quantization, one needs to demonstrate that its expectation value in an appropriate semiclassical state recovers $C_{gr}[N]$. A natural Gaussian coherent state in the algebraic dual space of some dense subset in $\mathcal{H}_{gr}$ reads \cite{AFJ}:
\begin{equation}\label{Gaussian CS1}
\begin{split}
	(\Psi_{\zeta}|\coloneq&\sum_{\lambda,\xi\in\mathbb{R}}e^{-\frac{\epsilon^{2}}{2}(\lambda-\lambda_{0})^2}e^{i\frac{\bar{A}_{1}|_{0}}{2}(\lambda-\lambda_{0})}\\
	\times&e^{-\frac{\omega^{2}}{2}(\xi-\xi_{0})^2}e^{i\frac{\bar{A}_{y}|_{0}}{2}(\xi-\xi_{0})}(\lambda,\xi|,
\end{split}
\end{equation}
where $\zeta\equiv \bigl(\epsilon,\bar{A}_{1}|_{0},\bar{\pi}^{1}|_{0}\equiv\hbar\beta\kappa\lambda_{0}/6,\omega,\bar{A}_{y}|_{0},\\\pi^{y}|_{0}\equiv\hbar\beta\kappa\xi_{0}/2\bigr)$ with the subscript 0 denoting a specific point in the phase space, $\epsilon$ and $\omega$ are the Gaussian spreads. For practical calculations, we also use the shadow of $(\Psi_{\zeta}|$ on a regular lattice given by
\begin{equation}\label{Psi}
\begin{split}
	|\Psi_{\zeta}\rangle\coloneq&\sum_{n,m\in\mathbb{Z}}e^{-\frac{\epsilon^{2}}{2}(n-\lambda_{0})^2}e^{-i\frac{\bar{A}_{1}|_{0}}{2}(n-\lambda_{0})}\\
	\times&e^{-\frac{\omega^{2}}{2}(m\mu_{y}-\xi_{0})^2}e^{-i\frac{\bar{A}_{y}|_{0}}{2}(m\mu_{y}-\xi_{0})}|n,m\mu_{y}\rangle.
\end{split}
\end{equation}
As discussed in \cite{Ashtekar:2003hd,Taveras,Willis:2004br}, $\zeta$ has to satisfy the following conditions of large volume and late time: $(\lambda_{0})^{-1}\!\ll\!\epsilon\!\ll\!|\bar{A}_{1}|_{0}|\!\ll\!1$ and $(\xi_{0})^{-1}\!\ll\omega\!\ll\!|\bar{A}_{y}|_{0}|\!\ll\!1$, such that $\left|\Psi_{\zeta}\right>$ is sharply peaked at $(\bar{A}_{1}|_{0},\bar{\pi}^{1}|_{0},\bar{A}_{y}|_{0},\pi^{y}|_{0})$ and the fluctuations are within specified tolerance. This can be verified by directly computing the expectation values and uncertainties of the fundamental operators as:
\begin{align}
	\langle\hat{\bar{A}}_{1}\rangle\!&=\!2\!\sin(\!\frac{\bar{A}_{1}|_{0}}{2}\!)\!e^{-\frac{\epsilon^2}{4}},&
	\langle\hat{\bar{\pi}}^{1}\rangle\!&=\!\frac{\hbar\beta\kappa}{6}\!\lambda_{0},\nonumber\\
	\langle\hat{\bar{A}}_{y}\rangle\!&=\!\frac{2}{\mu_{y}}\!\sin(\!\frac{\mu_{y}\bar{A}_{y}|_{0}}{2}\!)\!e^{-\frac{\bar{\omega}^2}{4}},&
	\langle\hat{\pi}^{y}\rangle\!&=\!\frac{\hbar\beta\kappa}{2}\!\xi_{0},\nonumber\\ 
	\Delta\hat{\bar{A}}_{1}\!&=\!\sqrt{2}\epsilon\!\cos(\!\frac{\bar{A}_{1}|_{0}}{2}\!),&
	\Delta\hat{\bar{\pi}}^{1}\!&=\!\frac{\hbar\beta\kappa}{6}\!\frac{1}{\sqrt{2}\epsilon},\nonumber\\
	\Delta\hat{\bar{A}}_{y}\!&=\!\sqrt{2}\omega\!\cos(\!\frac{\mu_{y}\bar{A}_{y}|_{0}}{2}\!),&
	\Delta\hat{\pi}^{y}\!&=\!\frac{\hbar\beta\kappa}{2}\!\frac{1}{\sqrt{2}\omega},\nonumber
\end{align}
where $\hat{\bar{A}}_{1}\!\equiv\! (\widehat{e^{i\bar{A}_{1}/2}}\!-\!\widehat{e^{-i\bar{A}_{1}/2}})/i$, $\hat{\bar{A}}_{y}\!\equiv\!(\widehat{e^{i\mu_{y}\bar{A}_{y}/2}}\!-\!\widehat{e^{-i\mu_{y}\bar{A}_{y}/2}})/(i\mu_{y})$, and we neglected the terms of order $\mathcal{O}(\epsilon^{2})$, $\mathcal{O}(\omega^{2})$, $\mathcal{O}(e^{-\pi^{2}/\epsilon^2})$, and  $\mathcal{O}(e^{-\pi^{2}/\bar{\omega}^2})$ with $\bar{\omega}\!\equiv\!\mu_{y}\omega$.

To compute the expectation value $\langle\hat{C}^{sym}_{gr}[N]\rangle$, one needs to evaluate $\langle\hat{B}^{sym}\rangle$, where $\hat{B}^{sym}\!\equiv\!(\hat{B}\!+\!\hat{B}^{\dagger})/2$ and $\hat{B}\!\in\!\{\hat{u}_{r},\hat{u}_{k,l}\}$. Since removing the absolute value contained in the non-analytic function $\eta(\lambda)$ introduces only a difference of order $\mathcal{O}(e^{-\epsilon^{2}\lambda_{0}^2})$ in the expectation value \cite{Ashtekar:2003hd,SFXY,Willis:2004br}, we can apply the Poisson summation formula together with the method of steepest descent to obtain (see Appendix \ref{appendixB} for details):
\begin{align}
\sum_{r=0,\pm4}\!\langle\hat{u}^{sym}_{r}\rangle\!&=\!\frac{4}{|\lambda_{0}|}\lambda_{0}^{2}[(1\!-\!2\sin^{2}(\bar{A}_{1}\!|_{0}))e^{-\!4\epsilon^2}\!-\!1\nonumber\\
&+\!\mathcal{O}(\frac{1}{\lambda_{0}^{2}},\frac{\epsilon^{-2}}{\lambda_{0}^{4}})]\!+\!\mathcal{O}(e^{-\frac{\pi^{2}}{\epsilon^2}},e^{-\epsilon^{2}\lambda_{0}^2}),\nonumber\\
\sum_{k,l=\pm2}\!\langle\hat{u}^{sym}_{k,l}\rangle\!&=\!-\frac{12}{|\lambda_{0}|}\!\lambda_{0}\xi_{0}[\sin(\bar{A}_{1}\!|_{0})\!\frac{\sin(\mu_{y}\bar{A}_{y}\!|_{0})}{\mu_{y}}\nonumber\\
&\times\!e^{-\!\epsilon^{2}-\!\bar{\omega}^{2}}\!+\!\mathcal{O}(\frac{1}{\lambda_{0}^{2}},\frac{1}{\lambda_{0}\xi_{0}},\frac{\epsilon^{-2}}{\lambda_{0}^{4}},\frac{\epsilon^{-2}}{\lambda_{0}^{3}\xi_{0}})]\nonumber\\
&+\mathcal{O}(e^{-\frac{\pi^{2}}{\bar{\omega}^2}},e^{-\frac{\pi^{2}}{\epsilon^2}},e^{-\epsilon^{2}\lambda_{0}^2}).\nonumber
\end{align}
Collecting these results, the expectation value up to the subleading order $\epsilon^{2}$ can be derived as:
\begin{equation}\label{expectation C}
\begin{split}
\langle\hat{C}^{sym}_{gr}[N]\rangle\!=&\frac{-3N\Delta^{-\frac{1}{2}}}{2\beta^{2}\kappa|\bar{\pi}^{1}\!|_{0}|}\![2(2\epsilon^{2}\!+\!\sin^{2}(\bar{A}_1\!|_{0}))(\bar{\pi}^{1}\!|_{0})^{2}\\
+&\!\sin(\bar{A}_1\!|_{0})\!\frac{\sin(\mu_{y}\bar{A}_{y}\!|_{0})}{\mu_{y}}\!\bar{\pi}^{1}\!|_{0}\pi^{y}\!|_{0}].
\end{split}
\end{equation}
Using the conditions satisfied by $\zeta$, $\langle\hat{C}^{sym}_{gr}[N]\rangle$ returns to $C_{gr}[N]$ in classical limit and hence is indeed a viable quantization. As a result \cite{Ding:2008tq,Yang:2009fp,Zhang:2021zfp}, the Eq.(\ref{expectation C}) with the subscript 0 omitted is regarded as the effective scalar constraint $\mathcal{C}_{gr}[N]$ for the geometric part. Note that the same effective scalar constraint can also be obtained through path
integral quantization in coherent state representations \cite{SFXY}. Note also that the higher-order corrections, such as $\lambda_{0}^{-2}$ and $\epsilon^{-2}\lambda_{0}^{-4}$, are not included into the effective scalar constraint. Actually, by including those terms, the effective scalar constraint would no longer be scaled by the same proportion under a scale transformation of the fiducial cell, and hence the resulting dynamics would depend on the artificially chosen auxiliary structure and thus lack predictive power. Specifically, under a rescaling of the fiducial cell, $\bar{A}_{1}|_{0}$ remains invariant, while $\bar{\pi}^{1}|_{0}$ is scaled by the same proportion. Thus, even at the level of fundamental operators, the expectation values $\langle\widehat{\sin(\bar{A}_1)}\!{}^{2}\rangle\!=\!2\epsilon^{2}\!+\!\sin^{2}(\bar{A}_1|_{0})$ and $\langle(\hat{\bar{\pi}}^{1})^{2}\rangle\!=\!(\hbar\beta\kappa/6)^{2}(2\epsilon^{2})^{-1}+(\bar{\pi}^{1}|_{0})^{2}$ involving $\epsilon$ cannot simultaneously transform in the same way as $\sin^{2}(\bar{A}_1|_{0})$ and $(\bar{\pi}^{1}|_{0})^{2}$ respectively. This reveals a limitation of the semiclassical approximation in preserving covariance.

For the matter sector, we can perform polymer-like quantization of the configuration variable $T$ in a fully parallel manner. It turns out that with a coherent state one can obtain the expectation value $\langle\hat{P}_{T}\rangle=P_{T}|_{0}$ up to order $\mathcal{O}(e^{-\pi^{2}/\sigma^2})$, with $\sigma$ being the corresponding Gaussian spread. Therefore, the total effective scalar constraint with quantum corrections of leading order reads $\mathcal{C}_{eff}[N]=\mathcal{C}_{gr}[N]+NcP_{T}$.

\vspace{\baselineskip}
\noindent\textit{Analysis of the effective dynamics.} Eq.(\ref{expectation C}) implies that the behavior of the Gaussian spread $\epsilon$ of the geometrical part is relevant for the effective dynamics of the quantum system, which will be discussed in two scenarios.

In the first scenario, the Gaussian spread $\epsilon$ is assumed to be a constant $\epsilon_{0}$. By choosing $\bar{\pi}^{1}>0,\pi^{y}>0$, $N=c$, and introducing the corresponding time parameter $\tau$, the equations of motion generated by $\mathcal{C}_{eff}[N]$ admit the following explicit solutions (see Appendix \ref{appendixC} for details):
\begin{eqnarray}
	&{}&\pi^{y}=\frac{\mu_{y}\bar{O}}{\sin(\mu_{y}\bar{A}_{y})},\label{piyva}\\
	&{}&\tan(\frac{\mu_{y}\bar{A}_{y}}{2})=\bar{F}[\frac{d_{0}-\cos(\bar{A}_{1})}{d_{0}+\cos(\bar{A}_{1})}]^{\frac{3}{4d_{0}}},\\
	&{}&\bar{\pi}^{1}=\frac{(\kappa\beta^{2}\Delta^{\frac{1}{2}}cP_{T}/3)-(\bar{O}\sin(\bar{A}_{1})/2)}{2\epsilon^{2}_{0}+\sin^{2}(\bar{A}_{1})},\label{pi1va}
\end{eqnarray}
where $P_T$, $d_{0}\!\equiv\!\sqrt{1+2\epsilon_{0}^{2}}$, $\bar{O}\!\equiv\!\pi^{y}\sin(\mu_{y}\bar{A}_{y})/\mu_{y}$ and $\bar{F}\!\equiv\!\tan(\mu_{y}\bar{A}_{y}/2)[(d_{0}+\cos(\bar{A}_{1}))/(d_{0}-\cos(\bar{A}_{1}))]^{3/(4d_{0})}$ are constants of motion. In the case of $\epsilon_{0}=0$, in the region of $\sin(\bar{A}_{1})>0$, we obtain
\begin{align}
\tan(\frac{\bar{A}_{1}}{2})\!=\!-\frac{c\tau}{\beta\Delta^{\frac{1}{2}}}+\sqrt{(\frac{c\tau}{\beta\Delta^{\frac{1}{2}}})^{2}+1}\!=:\!f(c\tau),\nonumber
\end{align}
where the integration constant corresponding to time translation freedom has been fixed by setting $\tan(\bar{A}_{1}/2)\!|_{\tau=0}\!=\!1$. Under the initial conditions of  $\bar{\pi}^{1}|_{\tau=0}\!>\!0$ and $0\!\ne\!|\bar{F}|\!\ll\!1$, one can conclude from Eqs.(\ref{piyva}) and (\ref{pi1va}) that both $\pi^{y}$ and $\bar{\pi}^{1}$ undergo a bounce respectively, thereby naturally avoiding the classical singularities. Moreover, between the two bouncing points, the visible universe undergoes a super-inflationary phase, during which a sufficient number of e-folds can be achieved by tuning the value of $\bar{F}$ \cite{SFXY}. In the case of $\epsilon_{0}\!\ne\!0$, by fixing $\bar{A}_{1}\!|_{\tau=0}\!=\!0$, $\bar{A}_{1}$ can be expressed as
\begin{align}
\bar{A}_{1}=-\arctan\left(\frac{\sqrt{2}\epsilon_{0}}{d_{0}}\tan(\frac{\sqrt{2}\epsilon_{0}d_{0}}{\beta\Delta^{1/2}}c\tau)\right)\!-\!n\pi,\nonumber
\end{align}
for 
\begin{align}
-\frac{\pi}{2}+n\pi\le\frac{\sqrt{2}\epsilon_{0}d_{0}}{\beta\Delta^{1/2}}c\tau\le\frac{\pi}{2}+n\pi,\quad
	n\in\mathbb{Z},
\end{align}
where the range of the $\arctan$ function is $[-\pi/2,\pi/2]$, with the convention $\arctan(\pm\infty)=\pm\pi/2$. Thus, under the initial condition of  $(\kappa\beta^{2}\Delta^{\frac{1}{2}}cP_{T}/3)-(|\bar{O}|/2)>0$, Eqs.(\ref{piyva}) and (\ref{pi1va}) indicate that the universe avoids the classical singularities by quantum bounces and becomes cyclic.

In the second scenario, we consider the possibility that the Gaussian spread $\epsilon$ depends on the dynamical variables, which turns out to be more interesting. As discussed in \cite{Ding:2008tq,Taveras}, to ensure that the coherent states remain semiclassical, thereby guaranteeing the correct classical limit of the quantum system, $\epsilon$ can be appropriately chosen as a function on the phase space such that $(\lambda_{0})^{-1}\!\ll\!\epsilon\!\ll\!|\bar{A}_{1}|_{0}|\!\ll\!1$. Taking account of $\bar{\pi}^{1}|_{0}=\hbar\beta\kappa\lambda_{0}/6$, and $\sin(\bar{A}_{1}|_{0})\approx\bar{A}_{1}|_{0}$ for $|\bar{A}_{1}|_{0}|\!\ll\!1$, we are motivated to choose $\epsilon$ as a slightly generalized geometric mean $|\bar{O}(\epsilon_{1}\sin(\bar{A}_1)+\epsilon_{2})/\bar{\pi}^{1}|^{1/2}$ between $1/\lambda_{0}$ and $|\bar{A}_{1}|_{0}|$, where $\epsilon_{1}$ and $\epsilon_{2}$ are positive constants to be determined by observations. The Dirac observable $\bar{O}$ is introduced as a simple choice to ensure that the dynamics governed by $\mathcal{C}_{eff}[N]$ are independent of the choice of fiducial cell and coordinates. By selecting positive $\sin(\bar{A}_1),\bar{\pi}^{1},\pi^{y},\bar{O}$, and $P_{T}$, and setting $N=c$ with the corresponding time parameter $\tau$, the explicit solution to the equations of motion reads (see Appendix \ref{appendixC} for details):
\begin{eqnarray}
	&{}&\tan(\frac{\bar{A}_{1}}{2})\!=\!f(c\tau),\\
	&{}&\pi^{y}=\frac{\mu_{y}\bar{O}}{\sin(\mu_{y}\bar{A}_{y})},\label{piyds}\\
	&{}&\tan(\frac{\mu_{y}\bar{A}_{y}}{2})=\bar{E}e^{-3\nu_{2}\tau}\tan^{3\nu_{1}}(\frac{\bar{A}_{1}}{2}),\\
	&{}&\bar{\pi}^{1}\!=\!\frac{(\kappa\beta^{2}\Delta^{\frac{1}{2}}cP_{T}/3)\!-\!2\epsilon_{2}\bar{O}}{\sin^{2}(\bar{A}_{1})}\!-\!\frac{\nu_{1}\bar{O}}{\sin(\bar{A}_{1})},\label{pi1ds}
\end{eqnarray}
where $\nu_{1}\!\equiv\!(1\!+\!4\epsilon_{1})/2$, $\nu_{2}\!\equiv\!2\epsilon_{2}\beta^{-1}\Delta^{-1/2}c$, and $\bar{E}\!\equiv\!\tan(\mu_{y}\bar{A}_{y}/2)\tan^{-3\nu_{1}}(\bar{A}_{1}/2)\times\exp(3\epsilon_{2}(\tan^{-1}(\bar{A}_{1}/2)-\tan(\bar{A}_{1}/2)))$ is constant of motion.

Eqs.(\ref{piyds}) and (\ref{pi1ds}) imply that, as long as $\bar{\pi}^{1}|_{\tau=0}\!>\!0$, the physical volume $2\pi\Delta^{1/2}\bar{\pi}^{1}$ of $\Sigma^3_{(0)}\!\times\mathbb{S}^1$ bounces at $\tau=0$, and the physical volume $\pi^{y}$ of $\Sigma^3_{(0)}$ also experiences another bounce assumed to be at $\tau\!=\!\tau_{0}$. Thus, the classical singularities could also be naturally avoided. For simplicity, we restrict our attention to the regime $\bar{E}\ll1$, where one has $\tau_{0}<0$, and the Hubble parameter $H\equiv\partial_{\tau}a/a$ reaches its maximum near $\tau=0$. This indicates that the interval $\tau_{0}\!<\!\tau\!<\!0$ corresponds to the super-inflationary phase of the scale factor $a$, by examining $\partial_{\tau}H\!>\!0$ (see Appendix \ref{appendixD} for details). The e-folding number for this phase can be estimated as:
\begin{equation}
	\ln(\frac{a|_{\tau=0}}{a|_{\tau=\tau_{0}}})\approx\frac{1}{3}\ln(\frac{\bar{E}^{-1}}{2}).
\end{equation}
Thus we may choose $\bar{E}=e^{-3\times55}/2$ to obtain the desired $55$ e-folding number \cite{Planck:2018jri}.

When $\tau\gg0$, the scale factors and the 4D matter density asymptotically take the forms:
\begin{eqnarray}
a&=&a_{0}e^{\nu_{2}\tau}\tau^{\nu_{1}},\label{asymptoticallyA}\\
b&=&\chi_{2}\Delta^{\frac{3}{2}}-\chi_{3}\Delta^{\frac{3\nu_{1}}{2}},\label{asymptoticallyB}\\
\rho^{(4)}&\equiv&\frac{P_{T}}{\pi^{y}}=\frac{2\bar{E}P_{T}}{\mu_{y}\bar{O}}e^{-3\nu_{2}\tau}(\frac{\beta\Delta^{\frac{1}{2}}}{2c\tau})^{3\nu_{1}},\label{rho4}
\end{eqnarray}
where 
\begin{eqnarray}
&{}&a_{0}=(\frac{\mu_{y}\bar{O}}{2V_{0}\bar{E}})^{\frac{1}{3}}(\frac{\beta\Delta^{\frac{1}{2}}}{2c})^{-\nu_{1}},\quad
\chi_{2}=\frac{\rho^{(4)}c^{3}\tau^{2}}{6\pi\beta\hbar},\nonumber\\
&{}&\chi_{3}=\frac{\bar{E}}{\mu_{y}}e^{-3\nu_{2}\tau}(\frac{\beta\tau^{-1}}{2c})^{3\nu_{1}-1}(\nu_{2}\tau+\nu_{1}).\nonumber
\end{eqnarray}
Let the present universe be at $\tau=\tau_{1}$. By combining Eq.(\ref{asymptoticallyA}) with the current Hubble parameter $H|_{\tau=\tau_{1}}\!=\!H_{obs}\!\equiv\!67.66\,\text{km}\!\cdot\!\text{s}^{-1}\!\cdot\!\text{Mpc}^{-1}$ and deceleration parameter $(-a\partial_{\tau}^{2}a/(\partial_{\tau}a)^{2})|_{\tau=\tau_{1}}\!=\!q_{obs}\!\equiv\!-0.53335$ \cite{Planck:2018vyg}, we obtain $\tau_{1}\!=\!\sqrt{\nu_{1}/(q_{obs}+1)}/H_{obs}$ and $\nu_{2}\!=\!H_{obs}(1\!-\!\sqrt{\nu_{1}(q_{obs}+1)})$. Moreover, by requiring the existence of a decelerating expansion phase after the end of super-inflation, we infer that $\nu_{1}<1$. Then this phase occurs for $0<\tau<\tau_2\equiv(\sqrt{\nu_{1}}-\nu_{1})/\nu_{2}$. Since the effective 4D gravitational "constant" in 5D KK cosmology reads $G^{(4)}\!=G^{(5)}/(2\pi b)$ \cite{Qiang:2004gg}, Eq.(\ref{asymptoticallyB}) can be rewritten as $\Delta^{\frac{3}{2}(1-\nu_{1})}=\chi_{3}/(\chi_{2}-\chi_{1})$, with $\chi_{1}:=c^{3}/(32\pi^{2}\beta\hbar G^{(4)})$, which requires $(\chi_{2}\!-\!\chi_{1})|_{\tau=\tau_{1}}\!>\!0$. Taking account of  $\rho^{(4)}|_{\tau=\tau_{1}}=3H_{obs}^{2}\Omega_{m}/(8\pi G_{obs})$, where $G^{(4)}|_{\tau=\tau_{1}}\!=\!G_{obs}\!\equiv\!6.674\!\times\!10^{-11}\,\text{m}^{3}\cdot\text{kg}^{-1}\cdot\text{s}^{-2}$ and $\Omega_{m}=0.3111$ \cite{Planck:2018vyg}, we obtain $\nu_{1}\!>(q_{obs}\!+\!1)/(2\Omega_{m})\!=\!0.75$ (see Appendix \ref{appendixD} for details).

Collecting the above results, the range of $\nu_{1}$ reads $0.75<\nu_{1}<1$, which implies a very small $\nu_{2}=\mathcal{O}(10^{-19}\text{s}^{-1})$. Therefore, after the end of the super-inflation, the scale factor $a$ first undergoes a decelerating expansion dominated by the term $\tau^{\nu_{1}}$, and then goes into an accelerating expansion dominated by the term  $e^{\nu_{2}\tau}$. This suggests that dark energy could indeed originate from quantum fluctuations.

To further determine the parameters of the model, we assume that $\Delta$ is of the same order as the 2D minimal area in the 4D LQG theory \cite{A. Ashtekar and Y. Ma,Ashtekar:2004eh,Rovelli:2004tv,Thiemann:2007pyv,Han:2005km} and take $\beta=0.1969227$ \cite{Song:2022zit}. Then by solving the equation $\Delta^{\frac{3}{2}(1-\nu_{1})}=[\chi_{3}/(\chi_{2}-\chi_{1})]|_{\tau=\tau_{1}}$ with respect to $\nu_{1}$, we get an estimation of $\nu_{1}\!=\!0.75+\mathcal{O}(10^{-150})$ (see Appendix \ref{appendixD} for details), and hence $\nu_{2}\!=\!8.955\times10^{-19}\,\text{s}^{-1}$, $\tau_{0}\!=\!-8.188\!\times\!10^{-13}\,\text{s}$, $\tau_{1}\!=\!18.33\,\text{Gyr}$, and $\tau_{2}\!=\!4.108\,\text{Gyr}$. Additionally, using Eq.(\ref{rho4}), the 4D and 5D critical densities can be estimated as $\rho^{(4)}_{c}\!\equiv\!P_{T}/(\pi^{y}|_{\tau=\tau_{0}})\!=\!7.588\!\times\!10^{184}\,\text{kg}/\text{m}^{3}$ and $\rho^{(5)}_{c}\!\equiv\!P_{T}/(2\pi\Delta^{1/2}\bar{\pi}^{1}|_{\tau=0})\!=\!4.525\!\times\!10^{131}\,\text{kg}/\text{m}^{4}$, where $\rho^{(5)}_{c}$ is of the same order as the 5D Planck density $\rho_{\text{P}}=(16\pi\beta)^{5/3}c^{-1}\Delta^{-5/2}\hbar$.
\begin{figure}[h]
	\includegraphics[width=0.48\textwidth,angle=0]{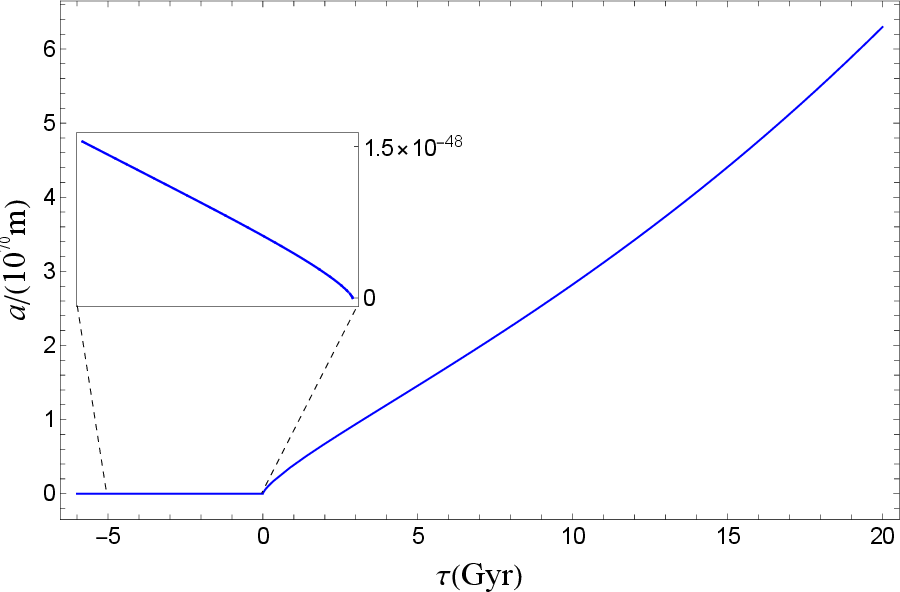}
	\caption{The evolution of the scale factor $a$, with the inset showing an enlarged view of the local region connected by dashed lines.}
	\label{Fig:a}
\end{figure}
\begin{figure}[H]
	\includegraphics[width=0.48\textwidth,angle=0]{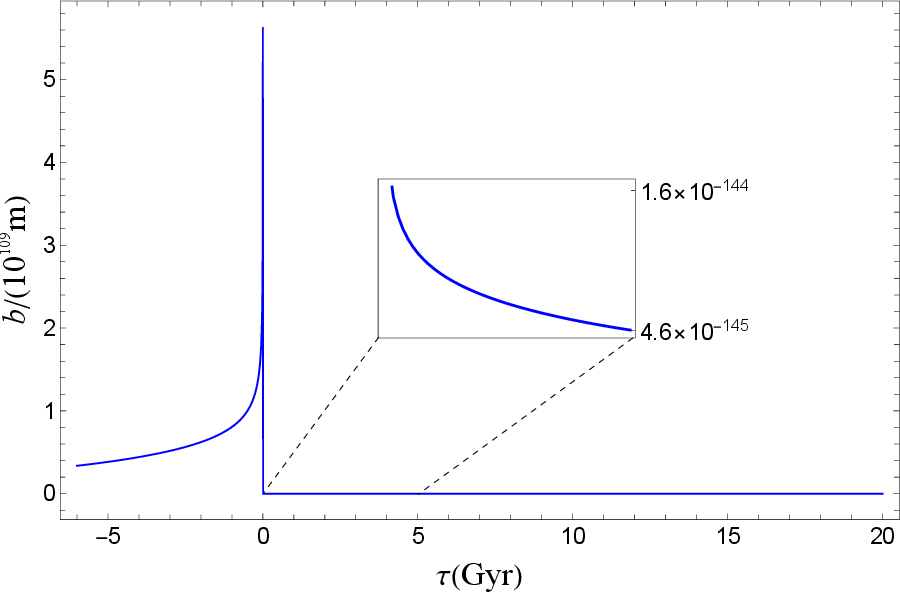}
	\caption{The evolution of the scale factor $b$, with the inset showing an enlarged view of the local region connected by dashed lines.}
	\label{Fig:b}
\end{figure}

With the above fixed parameters, the evolutions of the scale factors $a$ and $b$ are respectively shown in FIG.\ref{Fig:a} and FIG.\ref{Fig:b}, where we set $V_{0}\!=\!1$, $\bar{O}\!=\!1\,\text{m}^{3}$, and $P_{T}\!=\!\mu_{y}\bar{O}\rho^{(4)}_{c}$. It turns out that the expansion of $a$ after its bounce is accompanied by the collapse of $b$, followed by the asymptotic disappearance of $b$. This explains why the extra dimension is currently below the observational scale. Thus, our model provides a compactification mechanism for the extra dimension. Note that the chosen value of $\nu_{1}$ automatically ensures $G^{(4)}|_{\tau=\tau_{1}}\!=\!G_{obs}$, and that a decrease in $b$ implies an increase in $G^{(4)}$. However, a direct calculation using Eq.(\ref{asymptoticallyB}) with $G^{(4)}/G^{(4)}|_{\tau=\tau_{1}}=b|_{\tau=\tau_{1}}/b$ shows that $G^{(4)}$ increases by only $‌10\%$ over the next billion years. Therefore, the effective 4D gravitational "constant" $G^{(4)}$ may be approximately treated as a constant, in agreement with the observation. Note that, with the increasing precision of astronomical observations, it is possible to exam the time variation of $G^{(4)}$ in this model through gravitational wave signals in the future \cite{An:2023rqz,Sun:2023bvy,An:2025cmm}.

\vspace{\baselineskip}
\noindent\textit{Summary and discussion.} Our results in previous sections and the insights on them are summarized as follows: 
(i) The 5D KK LQC is set up based on the symmetric reduction of the connection formulation of the full theory, which opens a new avenue to understand the big issues of cosmology from fundamental physics. 
(ii) The effective scalar constraint of the 5D KK model, with not only the holonomy correction but also the first-order quantum fluctuation, is derived as a consistent result of the canonical and path integral approaches. The classical big bang and past big rip singularities are avoided by a quantum bounce and a quantum collapse respectively.
(iii) In contrast to 4D LQC models \cite{Ashtekar:2011rm}, our model can provide a super-inflationary phase capable of producing ‌55 e-folds‌ of expansion, which is sufficient to replace the conventional inflation mechanisms. This illustrates the origin of inflation as the effect of quantum gravity with extra dimensions.
(iv) The first-order quantum fluctuation of geometry may generate observable consequences at cosmological scales. In our second scenario, the effective dynamics of the KK LQC can archive both the sufficient super-inflation and the late-time accelerated expansion of the visible universe, and thus indicates a quantum geometric origin for dark energy. 
(v) In the current treatment, the higher-order quantum corrections beyond $\epsilon^2$ could not be included into the effective dynamics, ‌since they would lead to rescaling anomalies‌ of the fiducial cell and hence undermine the predictive power of the effective model.
 
Along with the amazing realization of inflation, late-time acceleration and dynamical compactification in our KK LQC model, there are several open issues. The radiative contributions in early-universe and corrections from full exponential terms of $e^{-\epsilon^{2}}$ and $e^{-\bar{\omega}^{2}}$ are not included in our treatments. In which regime and to what extent the neglected factors may alter predictions remains unclear. Also, a full quantum dynamical treatment-beyond semiclassical approximations is desirable to validate the proposed effective models. These issues are left for future studies.

\vspace{\baselineskip}
\noindent\textit{Acknowledgments.} This work is supported by National Natural Science Foundation of China (NSFC) with Grants No. 12275022 and No.12275087.

\onecolumngrid
\appendix

\section{Symmetry Reduction of Connection Variables}\label{appendixA}
In this section, we first summarize the fundamental results of an invariant connection theory obtained in Refs.\cite{HSV,Br,BK}, and then specialize these results to our particular framework.

Let $P(\Sigma,G)$ denote a principal fiber bundle over the base manifold $\Sigma$ with structure group $G$ equipped with a connection $\omega$. The symmetry group $\mathcal{S}$ acts on $\Sigma$ through the left action $\mathcal{L}:\mathcal{S}\times\Sigma\rightarrow\Sigma$. Given an open cover $\{U_{\alpha}\}$ of $\Sigma$ with corresponding local sections $\sigma_{\alpha}:U_{\alpha}\rightarrow P$, in the case that $\mathcal{L}(\mathcal{S},U_{\alpha})\subset U_{\alpha}$, the connection $A_{\alpha}:=\sigma_{\alpha}^*\omega$ is said to have the $\mathcal{S}$-symmetry if, for each $\alpha$, there exists a function $\rho_{\alpha}:\mathcal{S}\times U_{\alpha}\rightarrow G$ such that:
\begin{equation}\label{A symmetry condition}
	\mathcal{L}_s^*A_{\alpha}(x)=\rho_{\alpha}(s,x)A_{\alpha}(x)\rho_{\alpha}(s,x)^{-1}+\rho_{\alpha}(s,x)d\rho_{\alpha}(s,x)^{-1}, \quad\forall s\in\mathcal{S},\forall x\in U_{\alpha},
\end{equation}
where $\rho_{\alpha}$ are not arbitrary, but are constrained by the following two compatibility conditions:
\begin{equation}\label{rho condition}
	\begin{split}
		\rho_{\alpha}(s_1s_2,x)&=\rho_{\alpha}(s_2,x)\rho_{\alpha}(s_1,\mathcal{L}_{s_2}(x)),\\
		\rho_{\alpha}(s,x)k_{\alpha\beta}(\mathcal{L}_s(x))&=k_{\alpha\beta}(x)\rho_{\beta}(s,x),
	\end{split}
\end{equation}
where $k_{\alpha\beta}:U_{\alpha}\cap U_{\beta}\rightarrow G$ are the transition functions, i.e., $\sigma_{\beta}(x)=\sigma_{\alpha}(x)k_{\alpha\beta}(x)$. Condition (\ref{rho condition}) ensures that $A_{\alpha}$ retains $\mathcal{S}$-symmetry under  $\mathcal{S}$-symmetric and $G$-gauge transformations. We refer to $\rho_{\alpha}$, which satisfies condition (\ref{rho condition}), as the transformation function. Thus, to obtain the symmetry-reduced form of $A_{\alpha}$ one needs to first determine the transformation function and then solve Eq.(\ref{A symmetry condition}). However, a relative easy approach to derive the symmetry-reduced connection is to employ the fiber bundle theory, in which one obtains the following key theorem.

\begin{theorem}\label{theoremA1}
	Given the transformation function $\rho_{\alpha}$, $\forall\alpha$, there exists a unique left action $\tilde{\mathcal{L}}:\mathcal{S}\times P\rightarrow P$ on the principal bundle $P$, commuting with the right action of $G$, such that $\tilde{\mathcal{L}}$ induces the original left action $\mathcal{L}$ on $\Sigma$ and satisfies $\tilde{\mathcal{L}}_{s^{-1}}\circ\sigma_{\alpha}\circ\mathcal{L}_s(x)=\sigma_{\alpha}(x)\rho_{\alpha}(s,x)$. Furthermore, if $A_{\alpha}$ has $\mathcal{S}$-symmetry and $\rho_{\alpha}$ is its transformation function, then $\tilde{\mathcal{L}}_s^*\omega=\omega$. The converse of the above two statements also holds.
\end{theorem}

We call $P$ an $\mathcal{S}$-symmetric principal bundle if there exists a left action of the symmetry group $\mathcal{S}$ on $P$ that commutes with the right action of the gauge group $G$. \autoref{theoremA1} states that determining all possible $\rho_{\alpha}$ is equivalent to finding all inequivalent $\mathcal{S}$-symmetric principal bundles, where the bundle equivalence means that there exists a $\mathcal{S}$- and $G$- equivariant bundle isomorphism which induces the identity of the base manifold $\Sigma$. Moreover, solving the Eq.(\ref{A symmetry condition}) is equivalent to finding $\mathcal{S}$ invariant connection $\omega$. The following theorem identifies all $\mathcal{S}$-symmetric principal bundles.
\begin{theorem}\label{theoremA2}
	Fix a point $x\in\Sigma$, and let $\mathcal{F}$ be the isotropy subgroup of $\mathcal{S}$ at $x$. If both $\mathcal{F}$ and $G$ are compact, and all $\mathcal{S}$-orbits are of the same type, then an $\mathcal{S}$-symmetric principal bundle $P$ is uniquely characterized by a conjugacy class $[\phi]:=\{Ad_{g}\circ\phi\mid\forall g\in G\}$ of the homomorphism $\phi:\mathcal{F}\to G$, and a principal bundle $Q(\Sigma/\mathcal{S},Z_{G}(\phi(\mathcal{F})))$.
\end{theorem}

Here, the type of an orbit is the conjugacy class of the isotropy group at any point in the orbit, and $Z_{G}(\phi(\mathcal{F}))$ is the centralizer of $\phi(\mathcal{F})$ in $G$. The following theorem determines all $\mathcal{S}$-invariant connections on an $\mathcal{S}$-symmetric principal bundle:
\begin{theorem}\label{theoremA3}
	If $P$ is an $\mathcal{S}$-symmetric principal bundle labeled by $([\phi],Q)$ in \autoref{theoremA2}, then the $\mathcal{S}$-invariant connection $\omega$ on $P$ is uniquely determined by a connection $\tilde{\omega}$ on $Q$ and a Higgs field $\Phi:Q\to (L\mathcal{S})^*\otimes LG$, where $\Phi$ must satisfy the following conditions:
	\begin{equation}\label{Phi condition}
		\begin{split}
			\Phi_\mathfrak{q}(v_{\mathfrak{f}})&=\phi_*(v_{\mathfrak{f}}), \quad\forall \mathfrak{q}\in Q,\forall v_{\mathfrak{f}}\in L\mathcal{F},\\
			\Phi_\mathfrak{q}(Ad_\mathfrak{f}(v_{\mathfrak{s}}))&=Ad_{\phi(\mathfrak{f})}(\Phi_\mathfrak{q}(v_{\mathfrak{s}})), \quad\forall \mathfrak{q}\in Q,\forall \mathfrak{f}\in\mathcal{F},\forall v_{\mathfrak{s}}\in L\mathcal{S}.
		\end{split}
	\end{equation}
\end{theorem}

Here, the Higgs field $\Phi$ satisfies: $R^*_z\Phi=Ad_{z^{-1}}\Phi$, where $R:Q\times Z_{G}(\phi(\mathcal{F}))\to Q$ is the right action of the structure group $Z_{G}(\phi(\mathcal{F}))$ on $Q$. The following corollary provides the symmetric reduced form of $A_{\alpha}$.
\begin{corollary} 
	If $\omega$ is the $\mathcal{S}$-invariant connection labeled by $(\tilde{\omega},\Phi)$ in \autoref{theoremA3}, $\tilde{\sigma}_{\tilde{\alpha}}$ is the local section of $Q$, $\hat{\sigma}_{\hat{\alpha}}$ is the local section of the principal bundle $\mathcal{S}(\mathcal{S}/\mathcal{F},\mathcal{F})$, then there is a local section $\sigma_{\alpha}$ of $P$ such that:
	\begin{equation}\label{symmetric formA}
		A_{\alpha}=\tilde{\sigma}_{\tilde{\alpha}}^*\tilde{\omega}+(\tilde{\sigma}_{\tilde{\alpha}}^*\Phi)\circ(\hat{\sigma}_{\hat{\alpha}}^{-1}d\hat{\sigma}_{\hat{\alpha}}).
	\end{equation}
\end{corollary}

This corollary implies that any $A'_{\alpha}$ with $\mathcal{S}$-symmetry can be transformed into the form given by Eq.(\ref{symmetric formA}) via a $G$-gauge transformation. Hence, one may consider only connections of this form. Moreover, the dynamical information of $A_{\alpha}$ is captured by the connection $\tilde{\sigma}_{\tilde{\alpha}}^*\tilde{\omega}$ and the Higgs field $\tilde{\sigma}_{\tilde{\alpha}}^*\Phi$, both defined on $\Sigma/\mathcal{S}$. When a section different from $\tilde{\sigma}_{\tilde{\alpha}}$ is chosen, they undergo $Z_{G}(\phi(\mathcal{F}))$-gauge transformations in their respective ways. This explains the reduction of the gauge group from $G$ to $Z_{G}(\phi(\mathcal{F}))$.

In our case, the base manifold and the structure group read $\Sigma=\Sigma^4=\mathbb{R}^3\times\mathbb{S}^1$ and $G=\mathrm{Spin}(5)$. To be more general, we assume the symmetry group $\mathcal{S}=\mathbb{R}^3\rtimes\mathrm{SU}(2)$, which is the covering group of $\mathrm{E}(3)$. The homogeneity of spacetime along the extra dimensions, as discussed in the main text, can be simply imposed at the end. It is not difficult to see that the isotropic groups at each point in $\Sigma^4$ differ from each other by only adjoint transformations, so we can choose the isotropic group $\mathcal{F}=\{(0,u)\mid u\in\mathrm{SU}(2)\}=\mathrm{SU}(2)$.

We now focus our discussion on some $U_{\alpha}$, and omit the corner symbol $\alpha$ for simplicity. Similarly, we say that $\pi:=\pi^{aIJ}L_{IJ}\partial_a/2$ has $\mathcal{S}$-symmetry if and only if:
\begin{equation}\label{pi symmetry condition}
	\mathcal{L}_s^*\pi(x)=\rho(s,x)\pi(x)\rho(s,x)^{-1}, \quad\forall s\in\mathcal{S},\forall x\in U_{\alpha}.
\end{equation}
The key idea to find the symmetry-reduced form of $\pi$ is the following: if the $\mathcal{S}$-symmetric connection $A$ and $A'$ behave the same under symmetric transformations, then $A-A'$ and $\tilde{\pi}:=q_{ab}q^{-1/2}\pi^{bIJ}L_{IJ}dx^a/2$ also behave the same under symmetric transformations, where the spatial metric $q_{ab}$ is the inverse of $q^{ab}\equiv\pi^{aIJ}\pi^{b}{}_{IJ}/(2q)$, and $q\equiv\det(q_{ab})$. This implies that once the symmetry-reduced form of $A$ is determined, we automatically know the form of $A'$ and $\tilde{\pi}$, and thus the form of $\pi$. To determine the symmetry-reduced form of $A$, our first task is to find all conjugacy classes of group homomorphisms from $\mathrm{SU}(2)$ to $\mathrm{Spin}(5)$, namely $Hom(\mathrm{SU}(2),\mathrm{Spin}(5))/Ad$. To achieve this, we can apply the following theorem \cite{Varadarajan1984}.

\begin{theorem}\label{theoremA4}
	Let $LG_n(n=1,2)$ be the Lie algebra of the connected Lie group $G_n$, and $\varphi:LG_1\to LG_2$ be the Lie algebra homomorphism. If $G_1$ is simply connected, then there exists a unique Lie group homomorphism $\Upsilon:G_1\to G_2$ such that $\Upsilon_*=\varphi$, where $\Upsilon_*$ is the pushforward of $\Upsilon$ at the identity element of $G_1$.
\end{theorem}

This theorem implies that we can find all Lie group homomorphisms by solving the corresponding equations of the Lie algebra homomorphisms. By performing direct calculations, we obtain the following conclusion:
\begin{lemma}\label{LemmaA5}
	The conjugacy classes of group homomorphisms from $\mathrm{SU}(2)$ to $\mathrm{Spin}(5)$ can be expresed as $Hom(\mathrm{SU}(2),\mathrm{Spin}(5))/Ad=\{[\phi^0], [\phi^1], [\phi^2], [\phi^3]\}$, and their respective Lie algebra homomorphisms are:
	\begin{eqnarray}
		\phi^0_*(\tau_i)&=&0,\label{phi^0}\\
		\phi^1_*(\tau_i)&=&\tau_{i}^{+},\label{phi^1}\\
		\phi^2_*(\tau_3)&=&3\tau_{3}^{+}+\tau_{3}^{-}, \phi^2_*(\tau_1)=2\tau_{1}^{-}+\sqrt{3}L_{35}, \phi^2_*(\tau_2)=2\tau_{2}^{-}+\sqrt{3}L_{45},\label{phi^2}\\
		\phi^3_*(\tau_i)&=&\tau_{i}^{+}+\tau_{i}^{-},\label{phi^3}
	\end{eqnarray}
	where $\tau_j=-i\sigma_j/2$ are the generators of $\mathrm{SU}(2)$, $\sigma_j$ are the Pauli matrices, and $\tau_{i}^{\pm}:=\mp(L_{i4}\pm(\epsilon_i{}^{jk}L_{jk}/2))/2$.
\end{lemma}

Our second task is to compute the components of Eq.(\ref{symmetric formA}). Let the generators of the group $\mathcal{S}$ be $\{\hat{\tau}_{i},\hat{T}_{i}\}$, where $\hat{\tau}_{i}$ and $\hat{T}_{i}$ are the generators of the $\mathrm{SU}(2)$ subgroup and the $\mathbb{R}^3$ subgroup, respectively. These generators satisfy the relation:  $[\hat{\tau}_{i},\hat{\tau}_{j}]=\epsilon_{ij}{}^k\hat{\tau}_{k}$, $[\hat{\tau}_{i},\hat{T}_{j}]=\epsilon_{ij}{}^k\hat{T}_{k}$, $[\hat{T}_{i},\hat{T}_{j}]=0$. For simplicity, we choose the section $\hat{\sigma}:(x^i)\mapsto(x^i,\mathbb{I})$, such that $\hat{\sigma}^{-1}d\hat{\sigma}=\hat{T}_{i}dx^i$\cite{Bojowald:1999tr}. According to Eq.(\ref{Phi condition}), we also need to find the Higgs field $\tilde{\Phi}:=\tilde{\sigma}_{\tilde{\alpha}}^*\Phi:\mathbb{S}^1\supset U_{\tilde{\alpha}}\to(L\mathcal{S})^*\otimes LG$ that satisfies the following conditions:
\begin{equation}\label{tildePhi condition}
	\begin{split}
		\tilde{\Phi}_{y}(\hat{\tau}_{i})&=\phi_*(\hat{\tau}_{i}),\\
		\tilde{\Phi}_{y}([\hat{\tau}_{i},\hat{T}_{j}])&=[\tilde{\Phi}_{y}(\hat{\tau}_{i}),\tilde{\Phi}_{y}(\hat{T}_{j})].
	\end{split}
\end{equation}
By analyzing $\phi$ in Lemma \ref{LemmaA5}, we find that $\phi^0$, $\phi^1$, and $\phi^2$ lead to degenerate metrics, while only $\phi^3$ is physically appropriate. The corresponding form of $\tilde{\Phi}$ reads:
\begin{equation}\label{tildePhi}
	\tilde{\Phi}_{y}(\hat{\tau}_{i})=\tau_{i}^{+}+\tau_{i}^{-},\qquad
	\tilde{\Phi}_{y}(\hat{T}_{i})=\tilde{\Phi}^1(y)L_{i5}+\tilde{\Phi}^2(y)(\epsilon_i{}^{jk}/2)L_{jk}+\tilde{\Phi}^3(y)L_{i4},
\end{equation}
where the function $\tilde{\Phi}^i$ depends only on $y$. Furthermore, the reduced gauge group reads $Z_{G}(\phi^3(\mathcal{F}))=\exp(span_{\mathbb{R}}\{L_{45}\})=\mathrm{U}(1)$. By combining Eq.(\ref{tildePhi}) and Eq.(\ref{symmetric formA}), we obtain the symmetry-reduced forms of $A$ and $\pi$ as:
\begin{eqnarray}
	A_{aIJ}(x^{b})L^{IJ}dx^a/2&=&(\tilde{A}_1(y)L_{i5}+\tilde{A}_2(y)(\epsilon_i{}^{jk}/2)L_{jk}+\tilde{A}_3(y)L_{i4})dx^i+\tilde{A}_y(y)L_{45}dy,\\
	\pi^{aIJ}(x^{b})L_{IJ}\partial_a/2&=&(\tilde{\pi}^1(y)L^{i5}+\tilde{\pi}^2(y)(\epsilon^i{}_{jk}/2)L^{jk}+\tilde{\pi}^3(y)L^{i4})\partial_i+\tilde{\pi}^y(y)L^{45}\partial_y.
\end{eqnarray}
Because the symmetry group $\mathrm{U}(1)$, which characterizes homogeneity along the extra dimension, acts freely with the trivial isotropy subgroup $\{1\}$, the symmetric reduction of the connection variables can be obtained by imposing the independence of all components on the coordinate $y$.

To avoid divergences in the integration during the symmetry reduction from the full theory to the homogeneous model, we fix a fiducial cell $\Sigma^3_{(0)}:=(0,V_0^{1/3})^3\subset\mathbb{R}^3$, and restrict all the $\mathbb{R}^3$ integrals in the full theory to $\Sigma^3_{(0)}$ for the reduction. Then the gravitational symplectic structure $\Omega_{gr}$ is given by:
\begin{equation}
	\begin{split}
		\Omega_{gr}&=\frac{1}{2\beta\kappa^{(5)}}\int_{\Sigma^3_{(0)}}d^3x\int_{\mathbb{S}^1}dy\delta\pi^{aIJ}(x^{b})\wedge\delta A_{aIJ}(x^{b})\\
		&=\frac{V_0}{\beta\kappa}[\delta\tilde{\pi}^y\wedge\delta\tilde{A}_{y}+3\delta\tilde{\pi}^1\wedge\delta\tilde{A}_{1}+3\delta\tilde{\pi}^2\wedge\delta\tilde{A}_{2}+3\delta\tilde{\pi}^{3}\wedge\delta\tilde{A}_{3}],
	\end{split}
\end{equation}
where $\kappa^{(5)}:=2\pi\kappa$. By restricting constraints to reduced variables, we can get reduced constraints in which the vector constraint is automatically satisfied, and the simplicity and Gauss constraints are respectively equivalent to the following constraints \cite{Bodendorfer:2011nv}:
\begin{equation}
	S^{ab}_{M}=0\Leftrightarrow\tilde{\pi}^2=0, \qquad G^{IJ}=0\Leftrightarrow\tilde{A}_{1}\tilde{\pi}^{3}-\tilde{A}_{3}\tilde{\pi}^{1}=0.
\end{equation}
To solve the above constraints, we gauge fix $\tilde{A}_{2}=\tilde{\pi}^{3}=0$ and then solve  $\tilde{\pi}^2=\tilde{A}_{3}=0$ from the corresponding constraints. Additionally, we require the basic variables to remain invariant under the stretching coordinate transformation $x^i\to\alpha x^i$. To this end, $V_0$ is absorbed into the basic variables as:
\begin{equation}
	\pi^y:=V_0\tilde{\pi}^y, \qquad A_y:=\tilde{A}_y, \qquad \pi^1:=V_0^{2/3}\tilde{\pi}^1, \qquad A_1:=V_0^{1/3}\tilde{A}_1.
\end{equation}
Therefore, the phase space of the symmetry reduction is given by the following nontrivial Poisson brackets:
\begin{equation}
	\{A_y,\pi^{y}\}=\beta\kappa,\qquad\{A_1,\pi^{1}\}=\beta\kappa/3.
\end{equation}
Moreover, the connection variables are reduced to:
\begin{eqnarray}
	A_{aIJ}\frac{L^{IJ}}{2}dx^a&=&\frac{A_{1}}{V_{0}^{\frac{1}{3}}}L_{i5}dx^i+A_{y}L_{45}dy,\\
	\pi^{aIJ}\frac{L_{IJ}}{2}\partial_a&=&\frac{\pi^{1}}{V_{0}^{\frac{2}{3}}}L^{i5}\partial_i+\frac{\pi^{y}}{V_{0}}L^{45}\partial_y.
\end{eqnarray}

\section{Derivation of the expectation values}\label{appendixB}
As shown in the main text, the symmetric version of the scalar constraint operator can be written as $\hat{C}^{sym}_{gr}[N]=4^{-2}N\hbar\beta^{-1}\Delta^{-1/2}(\sum_{r\in\{0,4,-4\}}\hat{u}^{sym}_{r}+\sum_{k,l\in\{2,-2\}}\hat{u}^{sym}_{k,l})$, where $\hat{B}^{sym}:=(\hat{B}+\hat{B}^{\dagger})/2$, $\hat{B}\in\{\hat{u}_{r}, \hat{u}_{k,l}\}$. The property of this operator implies that we only need to consider the expectation values of the operators in the following form:
\begin{equation}\label{B1}
	\hat{H}_{\mathfrak{k},\mathfrak{l}}\left|\lambda,\xi\right>:=H_{1}(\lambda)H_{y}(\xi)\left|\lambda+\mathfrak{k},\xi+\mathfrak{l}\mu_{y}\right>,\qquad \mathfrak{k},\mathfrak{l}\in\mathbb{Z},
\end{equation}
where we first assume that $H_{1}(\lambda)$ and $H_{y}(\xi)$ are analytic functions and then extended the result to the non-analytic functions in our case. For the coherent state defined by Eq.(10) in the main text, a direct calculation gives
\begin{equation}\label{C2}
	\begin{split}
		\langle\Psi_{\zeta}|\hat{H}_{\mathfrak{k},\mathfrak{l}}|\Psi_{\zeta}\rangle&=e^{i\frac{\bar{A}_{1}|_{0}}{2}\mathfrak{k}-\frac{\epsilon^2}{4}\mathfrak{k}^{2}}\sum_{n\in\mathbb{Z}}H_{1}(n)e^{-\epsilon^{2}(n-\lambda_{0}+\frac{\mathfrak{k}}{2})^{2}}\\
		&\times e^{i\mu_{y}\frac{\bar{A}_{y}|_{0}}{2}\mathfrak{l}-\frac{\bar{\omega}^2}{4}\mathfrak{l}^{2}}\sum_{m\in\mathbb{Z}}H_{y}(m\mu_{y})e^{-\bar{\omega}^{2}(m-\bar{\xi}_{0}+\frac{\mathfrak{l}}{2})^{2}},
	\end{split}
\end{equation}
where $\bar{\omega}:=\mu_{y}\omega$, $\bar{\xi}_{0}:=(\mu_{y})^{-1}\xi_{0}$. By applying the Poisson summation formula and the steepest descent approximation to the summation in Eq.(\ref{C2}), we obtain the following result, retained up to the subleading order \cite{Ashtekar:2003hd,Yang:2022aec}:
\begin{equation}\label{C4}
	\begin{split}
		\langle\hat{H}_{\mathfrak{k},\mathfrak{l}}\rangle=\frac{\langle\Psi_{\zeta}|\hat{H}_{\mathfrak{k},\mathfrak{l}}|\Psi_{\zeta}\rangle}{\langle\Psi_{\zeta}|\Psi_{\zeta}\rangle}&=e^{i\frac{\bar{A}_{1}|_{0}}{2}\mathfrak{k}-\frac{\epsilon^2}{4}\mathfrak{k}^{2}}[H_{1}(\lambda_{0}-\frac{\mathfrak{k}}{2})+\frac{1}{4\epsilon^{2}}\partial^{2}_{\lambda}H_{1}(\lambda_{0}-\frac{\mathfrak{k}}{2})\\
		&+\mathcal{O}(\epsilon^{-4}\partial^{4}_{\lambda}H_{1}(\lambda_{0}-\frac{\mathfrak{k}}{2}))](1+\mathcal{O}(e^{-\frac{\pi^{2}}{\epsilon^2}})),\\
		&\times e^{i\mu_{y}\frac{\bar{A}_{y}|_{0}}{2}\mathfrak{l}-\frac{\bar{\omega}^2}{4}\mathfrak{l}^{2}}[H_{y}(\xi_{0}-\frac{\mathfrak{l}\mu_{y}}{2})+\frac{1}{4\omega^{2}}\partial^{2}_{\xi}H_{y}(\xi_{0}-\frac{\mathfrak{l}\mu_{y}}{2})\\
		&+\mathcal{O}(\omega^{-4}\partial^{4}_{\xi}H_{y}(\xi_{0}-\frac{\mathfrak{l}\mu_{y}}{2}))](1+\mathcal{O}(e^{-\frac{\pi^{2}}{\bar{\omega}^2}})).
	\end{split}
\end{equation}

Eq.(\ref{C4}) can be used to compute the expectation values of any operators in the form of Eq.(\ref{B1}). However, in the case of $\hat{u}_{r}$ and $\hat{u}_{k,l}$, the function $H_1$ will include $\eta(\lambda)\!=\!|\lambda|^{-1/2}$ if $\lambda\!\ne\!0$, which is non-analytic due to the absolute value. Fortunately, as shown below, this absolute value can be removed by certain technical treatments. Through direct computations, we get:
\begin{eqnarray}
	\left<\Psi_{\zeta}\left|\hat{u}_{0}\right|\Psi_{\zeta}\right>&=&-4\sum_{m\in\mathbb{Z}}e^{-\bar{\omega}^{2}(m-\bar{\xi}_{0})^{2}}\sum_{n\in\mathbb{Z}}\eta^{2}(n)n^{2}e^{-\epsilon^{2}(n-\lambda_{0})^{2}},\\
	\left<\Psi_{\zeta}\left|\hat{u}_{\pm 4}\right|\Psi_{\zeta}\right>&=&2e^{\pm i2\bar{A}_{1}|_{0}-4\epsilon^2}\sum_{m\in\mathbb{Z}}e^{-\bar{\omega}^{2}(m-\bar{\xi}_{0})^{2}}\sum_{n\in\mathbb{Z}}\eta(n\pm 4)\eta(n)n^{2}e^{-\epsilon^{2}(n-\lambda_{0}\pm 2)^{2}},\\
	\left<\Psi_{\zeta}\left|\hat{u}_{\pm 2,2}\right|\Psi_{\zeta}\right>&=&\pm 3(\mu_{y})^{-1}e^{\pm i\bar{A}_{1}|_{0}-\epsilon^2}e^{i\mu_{y}\bar{A}_{y}|_{0}-\bar{\omega}^2}\sum_{m\in\mathbb{Z}}m\mu_{y}e^{-\bar{\omega}^{2}(m-\bar{\xi}_{0}+1)^{2}}\nonumber\\
	&\times&\sum_{n\in\mathbb{Z}}\eta(n\pm 2)\eta(n)ne^{-\epsilon^{2}(n-\lambda_{0}\pm 1)^{2}},\\
	\left<\Psi_{\zeta}\left|\hat{u}_{\pm 2,-2}\right|\Psi_{\zeta}\right>&=&\mp 3(\mu_{y})^{-1}e^{\pm i\bar{A}_{1}|_{0}-\epsilon^2}e^{-i\mu_{y}\bar{A}_{y}|_{0}-\bar{\omega}^2}\sum_{m\in\mathbb{Z}}m\mu_{y}e^{-\bar{\omega}^{2}(m-\bar{\xi}_{0}-1)^{2}}\nonumber\\
	&\times&\sum_{n\in\mathbb{Z}}\eta(n\pm 2)\eta(n)ne^{-\epsilon^{2}(n-\lambda_{0}\pm 1)^{2}}.
\end{eqnarray}
Note that the function $\eta(\lambda)$ appears in the summations over $n$. These summations can be rewritten into the following two forms:
\begin{eqnarray}
	S_{\mathfrak{x}}:&=&\sum_{n\in\mathbb{Z}}\eta(n+\mathfrak{x})\eta(n-\mathfrak{x})(n+\mathfrak{x})^2e^{-\epsilon^{2}(n-\lambda_{0})^{2}},\qquad \mathfrak{x}=0,\pm 2,\label{B8}\\
	Q_{\mathfrak{y}}:&=&\sum_{n\in\mathbb{Z}}\eta(n+\mathfrak{y})\eta(n-\mathfrak{y})(n+\mathfrak{y})e^{-\epsilon^{2}(n-\lambda_{0})^{2}},\qquad \mathfrak{y}=\pm 1.\label{B9}
\end{eqnarray}
Then, we replace the function $\eta(\lambda)$ in Eqs.(\ref{B8}) and (\ref{B9}) by the complex function $\tilde{\eta}(\lambda)$, obtained by removing the absolute value, and denote the new summations as $\tilde{S}_{\mathfrak{x}}$ and $\tilde{Q}_{\mathfrak{y}}$. The new summations are different from the old ones by:
\begin{eqnarray}
	\delta S_{\mathfrak{x}}:=S_{\mathfrak{x}}-\tilde{S}_{\mathfrak{x}}&=&2\sum_{n\le-|\mathfrak{x}|-1}\eta(n+\mathfrak{x})\eta(n-\mathfrak{x})(n+\mathfrak{x})^2e^{-\epsilon^{2}(n-\lambda_{0})^{2}}\nonumber\\
	&+&(1-(-1)^{-1/2})\sum_{-|\mathfrak{x}|\le n\le|\mathfrak{x}|}\eta(n+\mathfrak{x})\eta(n-\mathfrak{x})(n+\mathfrak{x})^2e^{-\epsilon^{2}(n-\lambda_{0})^{2}},\label{B10}\\
	\delta Q_{\mathfrak{y}}:=Q_{\mathfrak{y}}-\tilde{Q}_{\mathfrak{y}}&=&2\sum_{n\le-|\mathfrak{y}|-1}\eta(n+\mathfrak{y})\eta(n-\mathfrak{y})(n+\mathfrak{y})e^{-\epsilon^{2}(n-\lambda_{0})^{2}}\nonumber\\
	&+&(1-(-1)^{-1/2})\sum_{-|\mathfrak{y}|\le n\le|\mathfrak{y}|}\eta(n+\mathfrak{y})\eta(n-\mathfrak{y})(n+\mathfrak{y})e^{-\epsilon^{2}(n-\lambda_{0})^{2}}.\label{B11}
\end{eqnarray}
Note that the second summations in both Eqs.(\ref{B10}) and (\ref{B11}) are finite, and the contributions from these terms are exponentially suppressed because of $\epsilon\lambda_{0}\gg 1$. Next, we consider the first terms in Eqs.(\ref{B10}) and (\ref{B11}). Their absolute values do not exceed $2\delta S'_{\mathfrak{r}}$ for $\mathfrak{r}=0,\pm1,\pm2$, where
\begin{equation}\label{B12}
	\begin{split}
		\delta S'_{\mathfrak{r}}:&=\sum_{n\le-|\mathfrak{r}|-1}\eta(n+\mathfrak{r})\eta(n-\mathfrak{r})(n+\mathfrak{r})^2e^{-\epsilon^{2}(n-\lambda_{0})^{2}}\\
		&=\sum_{n\ge 1}\eta(n+|\mathfrak{r}|-\mathfrak{r})\eta(n+|\mathfrak{r}|+\mathfrak{r})(n+|\mathfrak{r}|-\mathfrak{r})^2e^{-\epsilon^{2}(n+|\mathfrak{r}|+\lambda_{0})^{2}}\\
		&\le\sum_{n\ge 1}(n+|\mathfrak{r}|-\mathfrak{r})^2e^{-\epsilon^{2}(n+|\mathfrak{r}|+\lambda_{0})^{2}},
	\end{split}
\end{equation}
where, we used $\eta(\lambda)\le1$ for $\lambda\ge 1$. By applying the Euler-Maclaurin summation formula to Eq.(\ref{B12}), we obtain
\begin{equation}\label{B13}
	\delta S'_{\mathfrak{r}}\le\int_{1}^{\infty}F(x)dx+\frac{F(1)}{2}+\frac{1}{2}\int_{1}^{\infty}|\partial_{x}F(x)|dx,
\end{equation}
where $F(x):=h^{2}(x)e^{-\epsilon^{2}(h(x)+\mathfrak{r}+\lambda_{0})^{2}}$ with $h(x):=x+|\mathfrak{r}|-\mathfrak{r}$. It is not difficult to check that there exists at most only one point, denoted by $x_0$, in the interval $[1,\infty)$ such that $\partial_{x}F(x_0)=0$, where one has:
\begin{equation}
	h(x_{0})=\frac{\lambda_{0}+\mathfrak{r}}{2}(\sqrt{1+4\epsilon^{-2}(\lambda_{0}+\mathfrak{r})^{-2}}-1).
\end{equation} 
Thus, the last two term in the right hand side of Eq.(\ref{B13}) can be calculated as
\begin{equation}
	\frac{F(1)}{2}+\frac{1}{2}\int_{1}^{\infty}|\partial_{x}F(x)|dx=\begin{cases} F(x_{0}), &x_{0}>1,\\
		F(1), &x_{0}\le 1.\end{cases}
\end{equation}
It is easy to verify that both $F(1)$ and $F(x_{0})$ are exponentially suppressed by $e^{-\epsilon^{2}\lambda_{0}^2}$. The first term in the right hand side of Eq.(\ref{B13}) can be estimated by
\begin{equation}
	\begin{split}
		\int_{1}^{\infty}F(x)dx&=\int_{1+|\mathfrak{r}|-\mathfrak{r}}^{\infty}h^2e^{-\epsilon^{2}(h+\mathfrak{r}+\lambda_{0})^{2}}dh\\
		&<\int_{0}^{\infty}h^2e^{-\epsilon^{2}(h+\mathfrak{r}+\lambda_{0})^{2}}dh=\epsilon^{-3}\int_{0}^{\infty}h^2e^{-(h+\epsilon(\mathfrak{r}+\lambda_{0}))^{2}}dh\\
		&<\epsilon^{-3}\int_{0}^{\infty}e^{h^2-(h+\epsilon(\mathfrak{r}+\lambda_{0}))^{2}}dh=\frac{e^{-\epsilon^{2}(\mathfrak{r}+\lambda_{0})^{2}}}{2\epsilon^{4}(\mathfrak{r}+\lambda_{0})}.
	\end{split}
\end{equation}
Hence, Eq.(\ref{B13}) implies that $\delta S'_{\mathfrak{r}}$ is exponentially suppressed by $e^{-\epsilon^{2}\lambda_{0}^2}$. Therefore, the absolute values of $\hat{u}_{r}$, $\hat{u}_{k,l}$ can be removed in the calculation of their expectation values. Thus, we can safely use Eq.(\ref{C4}) to compute the following expectation values:
\begin{eqnarray}
	\langle\hat{u}^{sym}_{0}\rangle&=&-4(\lambda_{0})^{2}|\lambda_{0}|^{-1}[1+\mathcal{O}(e^{-\frac{\pi^{2}}{\epsilon^2}})+\mathcal{O}(e^{-\epsilon^2(\lambda_{0})^{2}})],\\
	\langle\hat{u}^{sym}_{4}+\hat{u}^{sym}_{-4}\rangle&=&(1-2\sin^{2}(\bar{A}_{1}|_{0}))e^{-4\epsilon^2}4(\lambda_{0})^{2}|\lambda_{0}|^{-1}[1+6(\lambda_{0})^{-2}+3(\lambda_{0})^{-4}\epsilon^{-2}\nonumber\\
	&+&\mathcal{O}((\lambda_{0})^{-4})+\mathcal{O}((\lambda_{0})^{-5}\epsilon^{-2})+\mathcal{O}(e^{-\frac{\pi^{2}}{\epsilon^2}})+\mathcal{O}(e^{-\epsilon^2(\lambda_{0})^{2}})],\\
	\langle\hat{u}^{sym}_{2,2}+\hat{u}^{sym}_{-2,-2}\rangle&=&\cos(\bar{A}_{1}|_{0}+\mu_{y}\bar{A}_{y}|_{0})e^{-\epsilon^{2}-\bar{\omega}^{2}}6(\mu_{y})^{-1}\lambda_{0}\xi_{0}|\lambda_{0}|^{-1}[1+\mu_{y}(\lambda_{0}\xi_{0})^{-1}+\frac{1}{2}(\lambda_{0})^{-2}\nonumber\\
	&+&\frac{\mu_{y}}{2\epsilon^{2}}(\lambda_{0})^{-3}(\xi_{0})^{-1}+\frac{3}{4\epsilon^{2}}(\lambda_{0})^{-4}+\mathcal{O}((\xi_{0})^{-1}(\lambda_{0})^{-3})+\mathcal{O}((\xi_{0})^{-1}(\lambda_{0})^{-4}\epsilon^{-2})\nonumber\\
	&+&\mathcal{O}((\lambda_{0})^{-4})+\mathcal{O}((\lambda_{0})^{-5}\epsilon^{-2})+\mathcal{O}(e^{-\frac{\pi^{2}}{\epsilon^2}})+\mathcal{O}(e^{-\frac{\pi^{2}}{\bar{\omega}^{2}}})+\mathcal{O}(e^{-\epsilon^2(\lambda_{0})^{2}})],\\
	\langle\hat{u}^{sym}_{-2,2}+\hat{u}^{sym}_{2,-2}\rangle&=&\cos(\bar{A}_{1}|_{0}-\mu_{y}\bar{A}_{y}|_{0})e^{-\epsilon^{2}-\bar{\omega}^{2}}6(-\mu_{y})^{-1}\lambda_{0}\xi_{0}|\lambda_{0}|^{-1}[1-\mu_{y}(\lambda_{0}\xi_{0})^{-1}+\frac{1}{2}(\lambda_{0})^{-2}\nonumber\\
	&-&\frac{\mu_{y}}{2\epsilon^{2}}(\lambda_{0})^{-3}(\xi_{0})^{-1}+\frac{3}{4\epsilon^{2}}(\lambda_{0})^{-4}+\mathcal{O}((\xi_{0})^{-1}(\lambda_{0})^{-3})+\mathcal{O}((\xi_{0})^{-1}(\lambda_{0})^{-4}\epsilon^{-2})\nonumber\\
	&+&\mathcal{O}((\lambda_{0})^{-4})+\mathcal{O}((\lambda_{0})^{-5}\epsilon^{-2})+\mathcal{O}(e^{-\frac{\pi^{2}}{\epsilon^2}})+\mathcal{O}(e^{-\frac{\pi^{2}}{\bar{\omega}^{2}}})+\mathcal{O}(e^{-\epsilon^2(\lambda_{0})^{2}})].
\end{eqnarray}

\section{Solving the Equations of Motion}\label{appendixC}

We consistently choose $\bar{\pi}^{1},\pi^{y}>0$, and set $N=c$ with the corresponding time parameter denoted by $\tau$. In the case where $\epsilon$ is a constant $\epsilon_{0}$, the equations of motion generated by $\mathcal{C}^{\epsilon_{0}}[N]\equiv\mathcal{C}_{eff}[N]|_{\epsilon=\epsilon_{0}}$ are:
\begin{eqnarray}
	\partial_{\tau}\bar{A}_{1}&=&\{\bar{A}_{1},\mathcal{C}^{\epsilon_{0}}[N]\}=-c\beta^{-1}\Delta^{-1/2}[2\epsilon_{0}^{2}+\sin^{2}(\bar{A}_{1})],\label{41.1}\\
	\partial_{\tau}\bar{A}_{y}&=&\{\bar{A}_{y},\mathcal{C}^{\epsilon_{0}}[N]\}=-c\frac{3}{2}\beta^{-1}\Delta^{-1/2}\sin(\bar{A}_{1})\frac{\sin(\mu_{y}\bar{A}_{y})}{\mu_{y}},\label{41.2}\\
	\partial_{\tau}\bar{\pi}^{1}&=&\{\bar{\pi}^{1},\mathcal{C}^{\epsilon_{0}}[N]\}=c\frac{1}{2}\beta^{-1}\Delta^{-1/2}\cos(\bar{A}_{1})(4\bar{\pi}^{1}\sin(\bar{A}_{1})+\pi^{y}\frac{\sin(\mu_{y}\bar{A}_{y})}{\mu_{y}}),\label{41.3}\\
	\partial_{\tau}\pi^{y}&=&\{\pi^{y},\mathcal{C}^{\epsilon_{0}}[N]\}=c\frac{3}{2}\beta^{-1}\Delta^{-1/2}\cos(\mu_{y}\bar{A}_{y})\pi^{y}\sin(\bar{A}_{1}).\label{41.4}
\end{eqnarray}
Dividing Eq.(\ref{41.2}) by Eq.(\ref{41.1}), we obtain:
\begin{equation}\label{41.5}
	\frac{\mu_{y}\partial_{\tau}\bar{A}_{y}}{\sin(\mu_{y}\bar{A}_{y})}=\frac{3}{2}\cdot\frac{\sin(\bar{A}_{1})\partial_{\tau}\bar{A}_{1}}{2\epsilon_{0}^{2}+\sin^{2}(\bar{A}_{1})}.
\end{equation}
Integrating both sides of Eq.(\ref{41.5}), we obtain:
\begin{equation}
	\partial_{\tau}\ln\left|\tan(\!\frac{\mu_{y}\bar{A}_{y}}{2}\!)\right|=\partial_{\tau}\ln(\!\Bigg(\!\frac{d_{0}\!-\!\cos(\bar{A}_{1})}{d_{0}\!+\!\cos(\bar{A}_{1})}\!\Bigg)^{\frac{3}{4d_{0}}}\!),
\end{equation}
where $d_{0}\!\equiv\!\sqrt{1+2\epsilon_{0}^{2}}$. This implies that $\bar{F}\!\equiv\!\tan(\mu_{y}\bar{A}_{y}/2)[(d_{0}+\cos(\bar{A}_{1}))/(d_{0}-\cos(\bar{A}_{1}))]^{3/(4d_{0})}$ is a Dirac observable, from which we obtain:
\begin{equation}
	\tan(\frac{\mu_{y}\bar{A}_{y}}{2})=\bar{F}[\frac{d_{0}-\cos(\bar{A}_{1})}{d_{0}+\cos(\bar{A}_{1})}]^{\frac{3}{4d_{0}}}.
\end{equation}
Based on the form of $\mathcal{C}^{\epsilon_{0}}[N]$, the Dirac observables $P_T$ and $\bar{O}\!\equiv\!\pi^{y}\sin(\mu_{y}\bar{A}_{y})/\mu_{y}$ can also be identified, which then yields: 
\begin{equation}\label{piys}
	\pi^{y}=\frac{\mu_{y}\bar{O}}{\sin(\mu_{y}\bar{A}_{y})}.
\end{equation}
Using the constraint $\mathcal{C}^{\epsilon_{0}}[N]=0$, we obtain:
\begin{equation}
	\bar{\pi}^{1}=\frac{(\kappa\beta^{2}\Delta^{\frac{1}{2}}cP_{T}/3)-(\bar{O}\sin(\bar{A}_{1})/2)}{2\epsilon^{2}_{0}+\sin^{2}(\bar{A}_{1})}.
\end{equation}
Therefore, once the evolution of $\bar{A}_{1}$ with respect to $\tau$ is known, the evolution of the remaining variables with respect to $\tau$ can be determined. In the case of $\epsilon_{0}=0$, in the region where $\sin(\bar{A}_{1})>0$, the integration of Eq.(\ref{41.1}) gives:
\begin{equation}\label{f1}
	\tan(\frac{\bar{A}_{1}}{2})=-\frac{c\tau}{\beta\Delta^{\frac{1}{2}}}+\sqrt{(\frac{c\tau}{\beta\Delta^{\frac{1}{2}}})^{2}+1}=:f(c\tau),
\end{equation}
where the integration constant corresponding to the time translation freedom has been fixed by setting $\tan(\bar{A}_{1}/2)\!|_{\tau=0}\!=\!1$. In the case of $\epsilon_0\ne0$, by fixing $\bar{A}_{1}\!|_{\tau=0}\!=\!0$ and integrating Eq.(\ref{41.1}), we obtain:
\begin{align}
	\bar{A}_{1}=-\arctan[\frac{\sqrt{2}\epsilon_{0}}{d_{0}}\tan(\frac{\sqrt{2}\epsilon_{0}d_{0}}{\beta\Delta^{1/2}}c\tau)]-n\pi,
\end{align}
for 
\begin{align}
	-\frac{\pi}{2}+n\pi\le\frac{\sqrt{2}\epsilon_{0}d_{0}}{\beta\Delta^{1/2}}c\tau\le\frac{\pi}{2}+n\pi,\quad
	n\in\mathbb{Z},
\end{align}
where the range of the arctan function is $[-\pi/2,\pi/2]$, with the convention $\arctan(\pm\infty)=\pm\pi/2$.

In the case where $\epsilon$ is the geometric mean $\bar{\epsilon}\equiv|\bar{O}(\epsilon_{1}\sin(\bar{A}_1)+\epsilon_{2})/\bar{\pi}^{1}|^{1/2}$, by choosing positive $\sin(\bar{A}_1),\bar{O}$ and $P_{T}$, the equations of motion generated by $\mathcal{C}^{\bar{\epsilon}}[N]\equiv\mathcal{C}_{eff}[N]|_{\epsilon=\bar{\epsilon}}$ are:
\begin{eqnarray}
	\partial_{\tau}\bar{A}_{1}&=&\{\bar{A}_{1},\mathcal{C}^{\bar{\epsilon}}[N]\}=-c\beta^{-1}\Delta^{-1/2}\sin^{2}(\bar{A}_{1}),\label{4.1}\\
	\partial_{\tau}\bar{A}_{y}&=&\{\bar{A}_{y},\mathcal{C}^{\bar{\epsilon}}[N]\}=-c\frac{3}{2}\beta^{-1}\Delta^{-1/2}[(1+4\epsilon_{1})\sin(\bar{A}_{1})+4\epsilon_{2}]\frac{\sin(\mu_{y}\bar{A}_{y})}{\mu_{y}},\label{4.2}\\
	\partial_{\tau}\bar{\pi}^{1}&=&\{\bar{\pi}^{1},\mathcal{C}^{\bar{\epsilon}}[N]\}=c\frac{1}{2}\beta^{-1}\Delta^{-1/2}\cos(\bar{A}_{1})(4\bar{\pi}^{1}\sin(\bar{A}_{1})+(1+4\epsilon_{1})\bar{O}),\label{4.3}\\
	\partial_{\tau}\pi^{y}&=&\{\pi^{y},\mathcal{C}^{\bar{\epsilon}}[N]\}=c\frac{3}{2}\beta^{-1}\Delta^{-1/2}\cos(\mu_{y}\bar{A}_{y})\pi^{y}[(1+4\epsilon_{1})\sin(\bar{A}_{1})+4\epsilon_{2}].\label{4.4}
\end{eqnarray}
Since Eq.(\ref{4.1}) coincides with Eq.(\ref{41.1}) for $\epsilon_{0}=0$, it follows that:
\begin{equation}\label{A1F}
	\tan(\frac{\bar{A}_{1}}{2})\!=\!f(c\tau).
\end{equation}
Combining Eq.(\ref{4.1}) and Eq.(\ref{4.2}), we obtain:
\begin{equation}\label{4.5}
	\frac{\mu_{y}\partial_{\tau}\bar{A}_{y}}{\sin(\mu_{y}\bar{A}_{y})}=3\nu_{1}\frac{\partial_{\tau}\bar{A}_{1}}{\sin(\bar{A}_{1})}-3\nu_{2},
\end{equation}
where $\nu_{1}\!\equiv\!(1\!+\!4\epsilon_{1})/2$ and $\nu_{2}\!\equiv\!2\epsilon_{2}\beta^{-1}\Delta^{-1/2}c$. Integrating both sides of Eq.(\ref{4.5}), we obtain:
\begin{equation}\label{D2}
	\partial_{\tau}\ln\left|\tan(\!\frac{\mu_{y}\bar{A}_{y}}{2}\!)\right|\!=\!\partial_{\tau}\ln(\!\left|\tan(\!\frac{\bar{A}_{1}}{2}\!)\right|^{3\nu_{1}}e^{-3\nu_{2}\tau}\!).
\end{equation}
By solving Eq.(\ref{A1F}) for $\tau$ as a function of $\bar{A}_{1}$ and substituting it into Eq.(\ref{D2}), one finds that $\bar{E}\!\equiv\!\tan(\mu_{y}\bar{A}_{y}/2)\tan^{-3\nu_{1}}(\bar{A}_{1}/2)\exp(3\epsilon_{2}(\tan^{-1}(\bar{A}_{1}/2)-\tan(\bar{A}_{1}/2)))$ is a Dirac observable, which gives:
\begin{equation}\label{Ay}
	\tan(\frac{\mu_{y}\bar{A}_{y}}{2})=\bar{E}e^{-3\nu_{2}\tau}\tan^{3\nu_{1}}(\frac{\bar{A}_{1}}{2}).
\end{equation}
Furthermore, the structure of $\mathcal{C}^{\bar{\epsilon}}[1]$ indicates that $P_T$ and $\bar{O}\!\equiv\!\pi^{y}\sin(\mu_{y}\bar{A}_{y})/\mu_{y}$ continue to be Dirac observables, so $\pi^y$ can still be expressed using Eq.(\ref{piys}). Applying the constraint $\mathcal{C}^{\bar{\epsilon}}[1]=0$ gives:
\begin{equation}\label{pi1s}
	\bar{\pi}^{1}=\frac{(\kappa\beta^{2}\Delta^{\frac{1}{2}}cP_{T}/3)-2\epsilon_{2}\bar{O}}{\sin^{2}(\bar{A}_{1})}-\frac{\nu_{1}\bar{O}}{\sin(\bar{A}_{1})}.
\end{equation}
Since the evolution of $\bar{A}_{1}$ with respect to $\tau$ is already given by Eq.(\ref{A1F}), the evolution of the remaining variables with respect to $\tau$ is fully determined.

\section{Observational Constraints on the Model Parameters}\label{appendixD}
In this section, we use observational data to constrain the parameters appearing in the case $\epsilon=|\bar{O}(\epsilon_{1}\sin(\bar{A}_1)+\epsilon_{2})/\bar{\pi}^{1}|^{1/2}$. Eq.(\ref{piys}) implies that $\pi^{y}$ experiences a bounce when $\sin(\mu_{y}\bar{A}_{y})=1$. Denoting the bounce point by $\tau\!=\!\tau_{0}$, it follows from Eqs.(\ref{Ay}) and (\ref{A1F}) that:
\begin{equation}\label{tau0}
	\bar{E}e^{-3\nu_{2}\tau_{0}}f^{3\nu_{1}}(c\tau_{0})=1.
\end{equation}
For simplicity, we restrict our attention to the regime $\bar{E}\ll1$, from which Eq.(\ref{tau0}) implies that $\tau_{0}<0$. From Eq.(\ref{4.4}), the Hubble parameter can be expressed as:
\begin{equation}\label{H}
	H\coloneq\frac{\partial_{\tau}a}{a}=\frac{c\cos(\mu_{y}\bar{A}_{y})}{2\beta\Delta^{\frac{1}{2}}}[(1+4\epsilon_{1})\sin(\bar{A}_{1})+4\epsilon_{2}].
\end{equation}
Because $\sin(\bar{A}_{1})|_{\tau=0}=1$ and
\begin{equation}
	\cos(\mu_{y}\bar{A}_{y})|_{\tau=0}=\left.\frac{1-\tan^{2}(\mu_{y}\bar{A}_{y}/2)}{1+\tan^{2}(\mu_{y}\bar{A}_{y}/2)}\right|_{\tau=0}=\frac{1-(\bar{E})^{2}}{1+(\bar{E})^{2}}\approx 1,
\end{equation}
the Hubble parameter $H$ reaches its maximum value near $\tau=0$. Furthermore, from Eqs.(\ref{4.1}), (\ref{4.2}), and (\ref{H}), it follows that:
\begin{align}
	\partial_{\tau}H&=3[\frac{1}{2}c\beta^{-1}\Delta^{-1/2}((1+4\epsilon_{1})\sin(\bar{A}_{1})+4\epsilon_{2})\sin(\mu_{y}\bar{A}_{y})]^2\nonumber\\
	&-\frac{1}{2}(1+4\epsilon_{1})(c\beta^{-1}\Delta^{-1/2}\sin(\bar{A}_{1}))^{2}\cos(\bar{A}_{1})\cos(\mu_{y}\bar{A}_{y}).\label{tH}
\end{align}
Using Eqs.(\ref{A1F}) and (\ref{Ay}), one can show that $\cos(\bar{A}_{1})\cos(\mu_{y}\bar{A}_{y})<0$ for $\tau_{0}<\tau<0$. It then follows from Eq.(\ref{tH}) that $\partial_{\tau}H>0$, which implies that the interval $\tau_{0}<\tau<0$ corresponds to the super-inflationary phase of the scale factor $a$. From Eqs.(\ref{piys}) and (\ref{A1F}), the e-folding number for this phase can be estimated as:
\begin{equation}
	\ln(\frac{a|_{\tau=0}}{a|_{\tau=\tau_{0}}})=\frac{1}{3}\ln(\frac{\bar{E}+\bar{E}^{-1}}{2})\approx\frac{1}{3}\ln(\frac{\bar{E}^{-1}}{2}).
\end{equation}
Thus we may choose $\bar{E}=e^{-3\times55}/2$ to obtain the desired $55$ e-folding number \cite{Planck:2018jri}.

Let the present universe be at $\tau=\tau_{1}$. By combining the late-time asymptotic expression of the visible-dimension scale factor, 
\begin{equation}\label{aasy}
	a=a_{0}e^{\nu_{2}\tau}\tau^{\nu_{1}},
\end{equation}
with the present values of the Hubble parameter $(\partial_{\tau}a/a)|_{\tau=\tau_{1}}\!=\!H_{obs}\!\equiv\!67.66\,\text{km}\!\cdot\!\text{s}^{-1}\!\cdot\!\text{Mpc}^{-1}$ and the deceleration parameter $(-a\partial_{\tau}^{2}a/(\partial_{\tau}a)^{2})|_{\tau=\tau_{1}}\!=\!q_{obs}\!\equiv\!-0.53335$ \cite{Planck:2018vyg}, we obtain
\begin{align}
	H_{obs}&=\nu_{2}+\nu_{1}(\tau_1)^{-1},\label{HO}\\
	q_{obs}&=\nu_{1}(H_{obs}\tau_1)^{-2}-1,
\end{align}
from which we can solve 
\begin{align}
	\tau_{1}&=\frac{1}{H_{obs}}\sqrt{\frac{\nu_{1}}{q_{obs}+1}},\label{T1}\\
	\nu_{2}&=H_{obs}(1-\sqrt{\nu_{1}(q_{obs}+1)})\label{nu2}.
\end{align}
Moreover, by requiring the existence of a decelerating expansion phase after the end of super-inflation and using Eq.(\ref{aasy}), we obtain in this phase:
\begin{equation}\label{qq}
	0<-\frac{a\partial_{\tau}^{2}a}{\partial_{\tau}a)^{2}}=\nu_{1}(\nu_{2}\tau+\nu_{1})^{-2}-1.
\end{equation}
From Eq.(\ref{qq}), it follows that this phase occurs for $0<\tau<\tau_2\equiv(\sqrt{\nu_{1}}-\nu_{1})/\nu_{2}$, with $\nu_{1}<1$. By combining the late-time asymptotic expression of the extra-dimensional scale factor, 
\begin{equation}
	b=\chi_{2}\Delta^{\frac{3}{2}}-\chi_{3}\Delta^{\frac{3\nu_{1}}{2}},
\end{equation}
with the effective 4D gravitational "constant" in 5D KK cosmology, 
\begin{equation}
	G^{(4)}=\frac{G^{(5)}}{2\pi b}=\frac{\Delta^{\frac{3}{2}}c^{3}}{32\pi^{2}\beta\hbar b},
\end{equation}
we obtain:
\begin{equation}\label{DD}
	\Delta^{\frac{3}{2}(1-\nu_{1})}=\frac{\chi_{3}}{\chi_{2}-\chi_{1}},
\end{equation}
where $\chi_{1}:=c^{3}/(32\pi^{2}\beta\hbar G^{(4)})$. From Eq.(\ref{DD}), since $\chi_{3}|_{\tau=\tau_{1}}>0$, it follows that 
\begin{equation}\label{X2X1}
	(\chi_{2}-\chi_{1})|_{\tau=\tau_{1}}>0.
\end{equation}
In addition, by combining Eq.(\ref{T1}), the definition of $\chi_{2}$, and $\rho^{(4)}|_{\tau=\tau_{1}}=3H_{obs}^{2}\Omega_{m}/(8\pi G_{obs})$, where $G^{(4)}|_{\tau=\tau_{1}}\!=\!G_{obs}\!\equiv\!6.674\!\times\!10^{-11}\,\text{m}^{3}\cdot\text{kg}^{-1}\cdot\text{s}^{-2}$ and $\Omega_{m}=0.3111$ \cite{Planck:2018vyg}, we obtain
\begin{equation}\label{X2}
	\chi_{2}|_{\tau=\tau_{1}}=\frac{\rho^{(4)}|_{\tau=\tau_{1}}c^{3}(\tau_{1})^{2}}{6\pi\beta\hbar}=\frac{\Omega_{m}c^{3}}{16\pi^{2}\beta\hbar G_{obs}}\frac{\nu_{1}}{q_{obs}+1}=\frac{2\Omega_{m}\nu_{1}}{q_{obs}+1}\chi_{1}|_{\tau=\tau_{1}}.
\end{equation}
Substituting Eq.(\ref{X2}) into Eq.(\ref{X2X1}) shows that $\nu_{1}\!>(q_{obs}\!+\!1)/(2\Omega_{m})\!=\!0.75$. Collecting the above results, the range of $\nu_{1}$ is $0.75<\nu_{1}<1$, from which Eq.(\ref{nu2}) implies $\nu_{2}=\mathcal{O}(10^{-19}\text{s}^{-1})$.

To further determine the parameters of the model, we assume that $\Delta$ is of the same order as the 2D minimal area $Ar^{(4)}_{min}=0.2375\times2\sqrt{3}\pi\hbar G_{obs}c^{-3}=6.7518\times10^{-70}\text{m}^{2}$ in the 4D LQG theory \cite{A. Ashtekar and Y. Ma,Ashtekar:2004eh,Rovelli:2004tv,Thiemann:2007pyv,Han:2005km} and take $\beta=0.1969227$ \cite{Song:2022zit}. To determine $\nu_{1}$, we need to solve the equation $\Delta^{\frac{3}{2}(1-\nu_{1})}=[\chi_{3}/(\chi_{2}-\chi_{1})]|_{\tau=\tau_{1}}$. Using Eqs.(\ref{X2}), (\ref{HO}), (\ref{T1}), and the definition of $\chi_{3}$, this equation can be rewritten as:
\begin{equation}\label{Snu2}
	\frac{2\Omega_{m}\nu_{1}}{q_{obs}+1}-1=\Delta^{-\frac{3}{2}(1-\nu_{1})}\left.\frac{\chi_{3}}{\chi_{1}}\right|_{\tau=\tau_{1}}=\Delta^{-\frac{3}{2}(1-\nu_{1})}\frac{\bar{E}}{\mu_{y}\chi_{1}|_{\tau=\tau_{1}}}e^{-3\nu_{2}\tau_{1}}(\frac{\beta(\tau_{1})^{-1}}{2c})^{3\nu_{1}-1}\sqrt{\frac{\nu_{1}}{1+q_{obs}}}.
\end{equation}
Substituting Eqs.(\ref{T1}) and (\ref{nu2}) into the right-hand side of Eq.(\ref{Snu2}), we denote the resulting function by $k(\nu_{1})$. Noting that the left-hand side of Eq.(\ref{Snu2}) is $4\nu_{1}/3-1$, the solution $\nu_1=:3/4+\delta\nu_{1}$ is required to satisfy:
\begin{equation}\label{kn}
	\delta\nu_{1}=\frac{3}{4}k(\frac{3}{4}+\delta\nu_{1}).
\end{equation}
Assuming that $\delta\nu_{1}$ is sufficiently small so that $k(3/4+\delta\nu_{1})\approx k(3/4)+\partial_{\nu_{1}}k(3/4)\delta\nu_{1}$, we can then solve Eq.(\ref{kn}) to obtain $\delta\nu_{1}=k(3/4)/[4/3-\partial_{\nu_{1}}k(3/4)]=4.3484\times10^{-150}$
, showing that the assumption is self-consistent. It follows that $\nu_{1}=0.75+\mathcal{O}(10^{-150})$. Inserting this result into Eqs.(\ref{tau0}), (\ref{T1}), (\ref{nu2}), and the definition of $\tau_{2}$ leads to $\nu_{2}\!=\!8.955\times10^{-19}\,\text{s}^{-1}$, $\tau_{0}\!=\!-8.188\!\times\!10^{-13}\,\text{s}$, $\tau_{1}\!=\!18.33\,\text{Gyr}$, and $\tau_{2}\!=\!4.108\,\text{Gyr}$. Moreover, from Eqs.(\ref{piys}) and (\ref{pi1s}), the 4D and 5D critical densities can be expressed as
\begin{align}
	\rho^{(4)}_{c}&\equiv\frac{P_{T}}{(\pi^{y}|_{\tau=\tau_{0}})}=\frac{P_{T}}{\mu_{y}\bar{O}},\\
	\rho^{(5)}_{c}&\equiv\frac{P_{T}}{2\pi\Delta^{1/2}\bar{\pi}^{1}|_{\tau=0}}=\frac{P_{T}/\bar{O}}{\kappa\beta^{2}\Delta^{\frac{1}{2}}c(P_{T}/\bar{O})/3-2\epsilon_{2}-\nu_{1}}.
\end{align}
By combining the late-time asymptotic expression of the 4D matter density, 
\begin{equation}
	\rho^{(4)}\equiv\frac{P_{T}}{\pi^{y}}=\frac{2\bar{E}P_{T}}{\mu_{y}\bar{O}}e^{-3\nu_{2}\tau}(\frac{\beta\Delta^{\frac{1}{2}}}{2c\tau})^{3\nu_{1}},
\end{equation}
with its present value $3H_{obs}^{2}\Omega_{m}/(8\pi G_{obs})$, we can infer the value of $P_{T}/\bar{O}$, allowing us to estimate $\rho^{(4)}_{c}\!=\!7.588\!\times\!10^{184}\,\text{kg}/\text{m}^{3}$ and $\rho^{(5)}_{c}\!=\!4.525\!\times\!10^{131}\,\text{kg}/\text{m}^{4}$.

\twocolumngrid

\end{document}